%% file: bispec.tex
\documentclass[prd,onecolumn,showpacs,nofootinbib]{revtex4}
\usepackage{times}
\usepackage{natbib}
\usepackage{epsfig}
\usepackage{color}
\usepackage{comment}
\usepackage{amsfonts}
\usepackage{amssymb}

\newcommand{\bdi}[1]{\hbox{\boldmath{$#1$}}}

\newcommand{\beeq}{\begin{equation}}
\newcommand{\eneq}{\end{equation}}
\newcommand{\bear}{\begin{eqnarray}}
\newcommand{\enar}{\end{eqnarray}}
\newcommand{\up}[1]{{\rm #1}}
\newcommand{\AVE}[1]{\langle#1\rangle}
\newcommand{\qqq}{{(q)}}

\newcommand{\dpnu}{\Delta\nu}    
\newcommand{\RA}{\rightarrow}
\newcommand{\LL}{\mathcal{L}}    
\newcommand{\Rbar}{{\bar R}}

\newcommand{\gbar}{\bar g}       
\renewcommand{\AA}{\mathcal{A}}  
\newcommand{\BB}{\mathcal{B}}    
\newcommand{\CC}{\mathcal{C}}    
\newcommand{\VCC}{C}             
\newcommand{\VBB}{B}             
\newcommand{\TCC}{C}             
\newcommand{\rbar}{\bar r}

\newcommand{\UU}{u}              
\newcommand{\VV}{\mathcal{U}}    
\newcommand{\dUU}{\delta\UU}     
\newcommand{\SVV}{v}             
\newcommand{\VVV}{v}             
\newcommand{\VP}{V_\parallel}   
\newcommand{\VTT}{V_\ttt}       
\newcommand{\VPP}{V_\pp}

\newcommand{\ax}{\alpha_{\chi}}     
\newcommand{\px}{\varphi_{\chi}}    
\newcommand{\pv}{\Psi}              
\newcommand{\deag}{{\delta e_\chi^\alpha}}      
\newcommand{\dnug}{\delta\nu_\chi}  
\newcommand{\dzg}{\dz_\chi}

\newcommand{\zz}{z}                 
\newcommand{\vhat}{\bdi{\hat v}}    
\newcommand{\ttt}{\theta}           
\newcommand{\pp}{\phi}              
\newcommand{\phat}{\hat\pp}         
\newcommand{\that}{\hat\ttt}        
\newcommand{\rhat}{\hat n}          
\newcommand{\nhat}{\bdi{\rhat}}     
\newcommand{\thatv}{\bdi{\that}}    
\newcommand{\phatv}{\bdi{\phat}}

\newcommand{\CG}{\hat g}         
\newcommand{\CK}{\hat k}         
\newcommand{\CU}{\hat u}         
\newcommand{\oo}{v}              
\newcommand{\cc}{\lambda}        
\newcommand{\dea}{{\delta e}}    
\newcommand{\eP}{\dea_\parallel} 
\newcommand{\dnu}{\delta\nu}     
\newcommand{\NN}{N}              
\newcommand{\NC}{\mathbb{C}}     
\newcommand{\DNC}{\delta\NC}     
\newcommand{\eTT}{\dea_\ttt}     
\newcommand{\ePP}{\dea_\pp}      
\newcommand{\dnn}{\delta n}      
\newcommand{\dhnu}{\widehat{\Delta\nu}}

\newcommand{\dTau}{\delta\mathcal{T}} 
\newcommand{\drr}{\delta r}     
\newcommand{\dtt}{\delta\ttt}   
\newcommand{\dpp}{\delta\pp}    

\newcommand{\dT}{\delta\tau}    
\newcommand{\dz}{\delta z}      
\newcommand{\DA}{\delta a}      
\newcommand{\HH}{\mathcal{H}}   
\newcommand{\Dcc}{\Delta\cc}    
\newcommand{\DT}{\Delta\tau}    
\newcommand{\DX}{\Delta x}

\newcommand{\dV}{\delta V}           
\newcommand{\dL}{\mathcal{D}_L}      
\newcommand{\ddL}{\delta\mathcal{D}_L} 
\newcommand{\dA}{\mathcal{D}_A}      
\newcommand{\DD}{\mathbb{D}}         
\newcommand{\dDD}{\delta\DD}         
\newcommand{\dg}{\delta g}           
\newcommand{\GG}{\delta\Gamma}

\begin{document}

\title{Beyond the linear-order relativistic effect in galaxy clustering:\\
Second-order gauge-invariant formalism}

\author{Jaiyul Yoo$^{1,2}$}
\author{Matias Zaldarriaga$^3$}
\affiliation{$^1$Institute for Theoretical Physics, University of Z\"urich,
CH-8057 Z\"urich, Switzerland}
\altaffiliation{jyoo@physik.uzh.ch}
\affiliation{$^2$Lawrence Berkeley National Laboratory, University of 
California, Berkeley, CA 94720, USA}
\altaffiliation{jyoo@lbl.gov}
\affiliation{$^3$School of Natural Sciences,
Institute for Advanced Study, Einstein Drive, Princeton, NJ 08540}

\received{25 April 2014}
\accepted{16 June 2014}

\begin{abstract}
We present the second-order general relativistic description of the observed
galaxy number density in a cosmological framework. The observed galaxy number
density is affected by the volume and the source effects, both of which arise
due to the mismatch between physical and observationally
inferred quantities such as the
redshift, the angular position, the volume, and 
the luminosity of the observed galaxies. These effects are 
computed to the second order in metric perturbations without choosing a gauge 
condition or adopting any restrictions on vector and tensor perturbations, 
extending the previous linear-order calculations.
Paying particular attention to the second-order gauge 
transformation, we explicitly isolate unphysical gauge modes and
construct second-order gauge-invariant variables. Moreover,
by constructing second-order tetrads in the observer's rest frame, 
we clarify the relation between the physical and the parametrized photon
wavevectors. Our second-order relativistic description will provide an
essential tool for going beyond the power spectrum in the era of precision
measurements of galaxy clustering. We discuss potential applications
and extensions of the second-order relativistic description of galaxy
clustering.
\end{abstract}


\maketitle

\section{Introduction}
Cosmology has seen its golden age, in particular due to the recent developments
in the cosmic microwave background experiments such as
the Wilkins Microwave Anisotropy Probe \cite{SPVEET03} and Planck 
\cite{PLANCK13} and in the large-scale galaxy surveys such as 
the Sloan Digital Sky Survey (SDSS; \cite{YOADET00}), the Two degree
Field Galaxy Redshift Survey (2dFGRS; \cite{CODAET01}),
and the Baryonic Oscillation Spectroscopic Survey (BOSS; \cite{SCBLET07}).
Furthermore, 
in order to exploit the enormous statistical power in three-dimensional 
volumes, a large number of galaxy surveys are planned to be operational
in a near future such as Euclid, the Dark Energy Spectroscopic Instrument 
(DESI), the Large Synoptic Survey Telescope, and the Wide-Field InfraRed Survey
Telescope (WFIRST), going progressively higher redshifts with larger
sky coverage. These surveys will be able to deliver precise measurements
of galaxy clustering on cosmological scales, in which alternative theories
of modified gravity or dark energy models deviate from general relativity
and in which the fingerprint of inflationary models remains intact.
In particular, this is the regime, in which the standard Newtonian description
of galaxy clustering breaks down, and therefore
it is crucial to have a proper
relativistic description to avoid misinterpretation of galaxy clustering
measurements on large scales.

The standard Newtonian description of galaxy clustering is based on
the assumption that the speed of light is infinite. However, 
the light we measure in galaxy surveys propagates throughout the Universe
at a finite speed,
and its path is affected, not only by the matter fluctuations,
but also by the relativistic contributions such as the gravitational potential
or the curvature of the Universe along its entire journey to reach us.
Therefore, the relation between the physical quantities of source
galaxies and the observable quantities in galaxy surveys is nontrivial,
and it requires a proper
relativistic treatment for solving the geodesic equation.
Given the observed redshift and the observed galaxy position on the sky,
 the full relativistic formula of galaxy clustering can be
derived \cite{YOFIZA09,YOO10}
by tracing the photon path backward in time and identifying the
relation of these observable quantities to the physical quantities of source
galaxies and the fluctuations that affect the photon path.

The relativistic formula provides the most accurate and complete description
of galaxy clustering on large scales and clarifies the physical origin
of all the effects in galaxy clustering \cite{YOO09}
such as the redshift-space distortion, the gravitational lensing, and
the Sachs-Wolfe effect. It was shown \cite{YOFIZA09,YOHAET12} that
the relativistic effect in galaxy clustering is measurable in the
current galaxy surveys and its detection significance can be greatly
enhanced if the multi-tracer technique \cite{SELJA09,SEHADE09} is employed,
which altogether provides great opportunities to test general relativity
and probe inflationary models on cosmological scales in upcoming
galaxy surveys. Furthermore, the relativistic
description of galaxy clustering has been independently verified in recent
years \cite{BODU11,CHLE11,JESCHI12,BRCRET12} and has received attention with
various applications (e.g., see \cite{MCDON09,BASEET11,CHZA11,SCJE12a,
JESC12,BEMAET12,LOMERI12,LOYOKO13,YOSE13a}).

The relativistic description of galaxy clustering is computed so far
to the linear order in metric perturbations. For Gaussian perturbations, the 
linear-order relativistic formula is all we need to describe galaxy clustering
on large scales, where perturbations are linear. However, the Universe
is far from being a complete Gaussian, even on large scales. For example,
many inflationary models have extra degrees of freedom supplied by
additional fields originating from the standard
particle physics models or its supersymmetric extensions
\cite{BAGR12,DIKAET09,BEMA09}. These new fields often couple to the 
inflaton field during the epoch of perturbation generation, and 
this nontrivial coupling leaves deviations
from statistical isotropy in the two-point correlation function of
curvature perturbation. Even in the absence
of additional fields in the simplest inflationary model, tensor modes of
gravitational waves can induce non-vanishing bispectrum in curvature
perturbations \cite{MALDA03}. 

Even on large scales the Universe deviates from
 perfect Gaussianity, and these inflationary
models manifest themselves in the three-point correlation function or 
the bispectrum in Fourier space. Since these three-point statistics
vanish at the linear order, the second-order relativistic effect is
needed to compute the three-point statistics and thereby to extract
additional information about non-standard inflationary models.
Furthermore, this non-trivial coupling is intrinsically subtle second-order
relativistic effect and affects not only the initial
curvature power spectrum, but also the photon propagation,
demanding consistent second-order
relativistic treatments for its observable effects in galaxy clustering. 
In other words, 
while the second-order relativistic effect is subtle and small in
comparison to the linear-order relativistic effect, it contains the distinct
physical information about the early Universe.
These arguments make a strong case for going beyond 
the linear-order relativistic effect in galaxy clustering. 
Here we develop the second-order general relativistic description of galaxy
clustering, providing an essential tool in the era of precision cosmology.

Compared to the linear-order calculations, the second-order calculation
is physically straightforward, albeit lengthy, but a few complications
arise due to nonlinearity inherent in beyond-the-leading-order calculations.
In particular, generation of vector and tensor perturbations is
inevitable, and their contributions to the observable quantities need to
be properly taken into consideration.
The organization of this paper is as follows. The main result is presented
in Sec.~\ref{sec:gcgr}, where we construct the full second-order relativistic
description of galaxy clustering. In Sec.~\ref{sec:gie}, we provide
the second-order gauge-invariant equations for those derived in
Sec.~\ref{sec:gcgr}, and we discuss the implications of our results 
in Sec.~\ref{sec:discussion}. The second-order gauge-invariant formalism
is presented in Appendix~\ref{app:gif}, and the relation between the photon
parametrization
and the observed angle is clarified in Appendix~\ref{app:photon}.
Throughout the paper we use Greek indices for 3D spatial components
and Latin indices for 4D spacetime components, respectively.
Various symbols are summarized in Table~\ref{tab:symbol}.

Further
in detail, the contents of Appendices and Sec.~\ref{sec:gcgr} are as follows.
In Appendix~\ref{app:con}, we present the general metric representation 
of a FRW universe and its decomposition into scalar, vector, and tensor.
In Appendix~\ref{app:gt}, the second-order gauge transformation is derived
and unphysical gauge mode is isolated. Finally, second-order gauge-invariant
variables are constructed in Appendix~\ref{app:gi}.
In Appendix~\ref{app:tetrads}, we construct the second-order tetrads,
representing the observer's rest frame in a FRW universe.
In Appendix~\ref{app:wavefrw}, the photon wavevector is constructed by using
local observable quantities, and the normalization of our photon 
parametrization is derived in Appendix~\ref{app:nor}. 
In Appendix~\ref{app:angle}, our choice for the normalization constant and
its relation to the observed angle is clarified.

Section~\ref{sec:gcgr} further divides into ten subsections. 
The first subsection provides an overview of the calculation in 
Section~\ref{sec:gcgr} and explains the physical origin of all the 
contributions to the observed galaxy fluctuation.
The parametrization of the photon wavevector
and the geodesic equation is presented in Sec.~\ref{ssec:geo}. The observed
redshift and the distortion in the observed redshift are derived 
in Sec.~\ref{ssec:obsz}. In Sec.~\ref{ssec:angle}, we briefly mention
the relation of our photon parametrization to the observed angle.
In Sec.~\ref{ssec:path}, the position of the source galaxies
is computed, and its deviation
from the inferred position is derived. In terms of the distortion in 
the observed redshift and the inferred source position, we present
the second-order deviations
in the observed solid angle in Sec.~\ref{ssec:kappa},
in the observed volume in Sec.~\ref{ssec:dV}, and 
in the luminosity distance in Sec.~\ref{ssec:dL}. Finally, the observed
galaxy number density to the second-order is derived in Sec.~\ref{ssec:ng},
and its fluctuation is presented in Sec.~\ref{ssec:delta}.

\section{Galaxy Clustering in General Relativity}
\label{sec:gcgr}

\subsection{Complete treatment of galaxy clustering:
Overview of the calculations}
\label{ssec:over}
Here we present a complete and unified treatment of galaxy clustering, 
providing an overview of the detailed calculations in Sec.~\ref{sec:gcgr}.
This treatment unifies all the effects in galaxy clustering
such as the redshift-space
distortion, the gravitational lensing, the Sachs-Wolfe effect, and their
relativistic effects into two physically distinct effects: the volume effect
and the source effect \cite{YOO09}. The volume effect describes the mismatch
between the physical volume occupied by the observed source galaxies and
the observationally inferred volume. The redshift-space distortion and the
gravitational lensing convergence, for example, arise from the volume effect. 
The source effect describes the contributions associated with the physical
properties of the source galaxy population. As the observationally inferred
properties of the source galaxies are different from their physical properties,
this mismatch gives rise to contributions to galaxy clustering. Magnification
bias is one example of the source effect.

In general, statistics in galaxy clustering are derived based on the
observed galaxy number density $n_g^\up{obs}$ (or the observed galaxy
fluctuation $\delta_g^\up{obs}$), and the observed galaxy number density
is further constructed using the basic observable quantities in galaxy
surveys: the observed redshift~$\zz$, the observed angular position 
$\nhat=(\ttt,\pp)$
of the source galaxy, and the number of galaxies $dN_g^\up{obs}$ counted
within the observed redshift and solid angle. The observationally inferred
volume $dV_\up{obs}$ occupied by the observed source galaxies is
\beeq
dV_\up{obs}={\rbar^2(\zz)\over H(\zz)(1+\zz)^3}~\sin\ttt~d\zz~d\ttt~d\pp~,
\eneq
where $\rbar$ is the comoving line-of-sight distance and $H$ is the Hubble
parameter, and the observed galaxy number density is then obtained as
\beeq
\label{eq:ngobs1}
n_g^\up{obs}(\zz,\nhat)={dN_g^\up{obs}(\zz,\nhat)\over dV_\up{obs}}~.
\eneq
Considering that the observed galaxies have the physical galaxy
number density $n_g$ and occupy the physical volume $dV_\up{phy}$, we can 
related the observed galaxy number density to those quantities as
\beeq
\label{eq:ngobs2}
dN_g^\up{obs}=n_g^\up{obs}dV_\up{obs}=n_gdV_\up{phy}~,\qquad
n_g^\up{obs}=n_g~(1+\dV)~,\qquad 
{dV_\up{phy}\over dV_\up{obs}}\equiv1+\dV~,
\eneq
where we defined the volume distortion $\dV$. Regardless of which the source
galaxy population $n_g$ is used, the volume distortion $\dV$ will always
contribute to galaxy clustering, and its contribution
is collectively described as the volume effect. In the following subsections,
we will compute the physical volume $dV_\up{phy}$ (and hence $\dV$)
and discuss the physical effects that contribute to the volume distortion.

The observed galaxies are grouped as galaxy samples based on various observable
quantities such as the rest-frame luminosity, the spectral color, 
and so on. However,
this classification is also based on the observationally inferred quantities,
and they differ from the physical quantities of the source galaxy population.
Therefore, when the source galaxy population is expressed in terms of 
observable quantities, this mismatch gives rise to contributions to
the observed galaxy number density $n_g^\up{obs}$ as for the volume
distortion~$\dV$. However, this source effect depends on which observable
quantities are used to define the galaxy sample and how the physical galaxy
number density $n_g$ representing the galaxy sample depends on the observable
quantities. Therefore, we will consider only the rest-frame luminosity,
the most frequently used quantity in galaxy surveys, 
to illustrate the source effect in Sec.~\ref{ssec:ng}.
In contrast, the time coordinate (or the distance
from us) of the observed galaxy sample is computed based on the observed
redshift, and hence the time evolution of the physical galaxy number density
will always contribute to galaxy clustering as one of the source effect
(see also Sec.~\ref{ssec:ng}).

\subsection{Photon geodesic equation}
\label{ssec:geo}
The photon path is described by a null geodesic $x^a(\oo)$ with an affine 
parameter~$\oo$, and its propagation direction is then $k^a(\oo)=dx^a/d\oo$ 
subject to the null condition $k^ak_a=0$ and the geodesic equation 
$k^a_{\;\;;b}k^b=0$.
We choose the affine parameter~$\oo$ such that 
the photon frequency measured in the rest frame of an observer with four 
velocity $u^a$ is 
\beeq
2\pi\nu=-g_{ab}k^a(\oo)u^b(\oo)~,
\label{eq:affineoo}
\eneq
where the four velocity of the observer is normalized as $u^au_a=-1$
(see Appendix~\ref{app:gif} for our notation convention).
Once the affine parameter is fixed in terms of physical quantities, the photon
wavevector $k^a(\oo)$ is completely set without any further degrees of freedom.
Since null geodesics are conformally 
invariant, we simplify the photon propagation equations by considering
a conformal transformation $ds^2=g_{ab}x^ax^b=a^2\CG_{ab}dx^adx^b$,
where the expansion factor~$a$ is removed in a conformally transformed
metric $\CG_{ab}$. Under the conformal transformation, the null geodesic
$x^a(\oo)$ remains unaffected, but its affine parameter is transformed to
another affine parameter~$\cc$ \cite{WALD84}:
\beeq
{d\oo\over d\cc}=\NC a^2~,~~~~~\CK^a={dx^a\over d\cc}
=\NC a^2 k^a~,~~~~~\CU^a=au^a~,
\label{eq:conf}
\eneq
where the unspecified proportionality constant~$\NC$ represents arbitrariness
or additional degree of freedom
in the conformally transformed affine parameter.

The conformally transformed wavevector~$\CK^a$ still satisfies
the same null condition $\CK^a\CK_a=0$ and the geodesic equation
$\CK^a_{\;\;;b}\CK^b=0$. While the physical wavevector~$k^a$ and its affine
parameter~$\oo$ are completely fixed, we have additional freedom to choose
its normalization and affine parameter~$\cc$ in the conformally transformed
metric. With this freedom, we parametrize the photon wavevector as
 (see Appendix~\ref{app:photon})
\beeq
\CK^a(\cc)=\left({d\tau\over d\cc},{dx^\alpha\over d\cc}\right)=
\bigg[1+\dnu,~-\left(e^\alpha+\dea^\alpha\right)\bigg]~,
\label{eq:wave}
\eneq
where $e^\alpha$ based on $\gbar_{\alpha\beta}$ 
is the photon propagation direction
normalized as $e^\alpha e_\alpha=1$ and the dimensionless quantities
$\dnu$ and $\dea^\alpha$ represent perturbations to the photon wavevector.
These perturbation variables are defined in a non-perturbative way, 
such that
they contain higher order perturbations, e.g., 
$\dnu=\dnu^{(1)}+\dnu^{(2)}+\cdots$ and by construction
$\AVE{\dnu}=0$ to all orders in perturbation.
Since the photon path is parametrized by the affine parameter~$\cc$, we have
\beeq
\label{eq:dlambda}
{d\over d\cc}={d x^a \over d\cc}
{\partial \over \partial x^a} =\CK^a \partial_a
=\left(\partial_\tau - e^\alpha \partial_\alpha\right)+\dnu~ \partial_\tau 
-\dea^\alpha \partial_\alpha~,
\eneq
where  the background relation in the round bracket
simply represents that the photon propagation 
path is a straight line to the zeroth order. However, to the second order
in perturbations, we need to consider the evolution of perturbations along
the photon path deviating from the straight line.

Using the conformally transformed metric, the photon wavevector can be
written to the second order in perturbations as
\bear
\CK_0&=&\CG_{0a}\CK^a=
-\left(1+\dnu+2\AA-\BB_\alpha e^\alpha+2~\AA~\dnu-\BB_\alpha
~\dea^\alpha\right)~, \nonumber \\
\CK_\alpha&=&\CG_{\alpha b}\CK^b=
-\left(e_\alpha+\dea_\alpha+\BB_\alpha+2~\CC_{\alpha\beta}
e^\beta+\dnu~\BB_\alpha+2~\CC_{\alpha\beta}~\dea^\beta\right)~,
\enar
and the null equation is then
\beeq
0=\CK^a\CK_a=\left(e^\alpha e_\alpha-1\right)
+2\left(e^\alpha\dea_\alpha-\dnu-\AA
+\BB_\alpha~e^\alpha+\CC_{\alpha\beta}~e^\alpha e^\beta\right)
-\dnu\bigg[\dnu+2~\left(2\AA-\BB_\alpha e^\alpha\right)\bigg]
+\bigg[\dea_\alpha
+2\left(\BB_\alpha+2~\CC_{\alpha\beta}~e^\beta\right)\bigg]\dea^\alpha~,
\label{eq:null}
\eneq
where the metric perturbations are defined in Appendix~\ref{app:gif}.
The background relation is trivially satisfied by the construction of the unit
direction vector $e^\alpha$.
Defining the perturbation to the observer four velocity,
we derive the four velocity vector of the observer
from the normalization condition $\UU^a\UU_a=-1$,
\bear
\label{eq:fourv}
\UU^\alpha&\equiv&{\VV^\alpha\over a}~,\qquad\qquad
\UU^0={1\over a}\left[1-\AA+{3\over 2} \AA^2 + {1\over 2}\VV^\alpha\VV_\alpha 
- \VV^\alpha \BB_\alpha\right]\equiv{1+\dUU^0\over a}~,\\
\UU_\alpha&=& g_{\alpha b}\UU^b=a\left(
\VV_\alpha-\BB_\alpha+\AA \BB_\alpha+2\VV^\beta \CC_{\alpha\beta}\right)
\equiv a\left(-\SVV_{,\alpha}+\VVV_\alpha\right)  ~,  \qquad
\UU_0 =g_{0a}\UU^a=-a\left( 1 + \AA - {1\over 2} \AA^2+ {1\over 2} \VV^\alpha 
\VV_\alpha \right)~. \nonumber 
\enar
To the second order in perturbation, the temporal component of the geodesic 
equation is
\beeq
\label{eq:tempo}
0=\CK^a\CK^0_{\;\; ;a}={d\over d\cc}\dnu+\hat\Gamma^0_{ab}\CK^a\CK^b=
\dnu^\prime-e^\alpha\dnu_{,\alpha}+\dnu\dnu^\prime-\dea^\alpha\dnu_{,\alpha}
+\GG^0~,\qquad
{d\over d\cc}\dnu=-\GG^0~,
\eneq
where $\hat\Gamma^a_{bc}$ is the Christoffel symbol based on $\CG_{ab}$,
the background relation for the temporal component is already removed 
by the construction of the conformally transformed
wavevector in Eq.~(\ref{eq:wave}), and we defined
\bear
\label{eq:GG0}
&&\hspace{-10pt}
\GG^0\equiv\hat\Gamma^0_{ab}\CK^a\CK^b=\AA'-2\AA_{,\alpha}e^\alpha
+\left(\BB_{\alpha|\beta}+\CC_{\alpha\beta}'\right) e^\alpha e^\beta
+2\dnu\left(\AA' - \AA_{,\alpha}e^\alpha\right)
-2\AA\AA' -\AA_{,\alpha}\BB^\alpha+\BB^\alpha\BB_\alpha' \\
&&\hspace{-10pt}
+2\left(2\AA\AA_{,\alpha}+\BB_\beta\CC_\alpha^{\beta\prime}
+\BB^\beta \BB_{[\alpha|\beta]} \right) e^\alpha
-2\dea^\alpha\left[\AA_{,\alpha}
-e^\beta\left(\BB_{(\alpha|\beta)} +\CC_{\alpha\beta}'\right)\right]
- \left[\AA\left( 2\BB_{\alpha|\beta} 
+2\CC_{\alpha\beta}'\right)+\BB_\gamma \left( 2 \CC^\gamma_{\alpha|\beta}
-\CC_{\alpha\beta}^{\;\;\;\;|\gamma}\right)\right]e^\alpha e^\beta
~. \nonumber
\enar
Similarly, the spatial component of the geodesic equation is
\bear
\label{eq:spatial}
0&=&\CK^b\CK^\alpha_{\;\; ;b}=-e^{\alpha\prime} + e^\beta e^\alpha_{\;\;,\beta}
-{d\over d\cc}\dea^\alpha+\hat\Gamma^\alpha_{bc}\CK^b\CK^c \\
&=&
\left(-e^{\alpha\prime} + e^\beta e^\alpha_{\;\;|\beta}\right)
-\dnu e^{\alpha\prime}+ \dea^\beta e^\alpha_{\;\;,\beta}
-\dea^{\alpha\prime} +e^\beta \dea^\alpha_{\;\;,\beta} 
-\dnu\dea^{\alpha\prime}+ \dea^\beta \dea^\alpha_{\;\;\;,\beta}+\GG^\alpha~,
\qquad {d\over d\cc}\dea^\alpha=\GG^\alpha~, \nonumber
\enar
where we defined a perturbation quantity in a similar way as
\bear
\label{eq:GGA}
&&\hspace{-10pt}
\GG^\alpha\equiv\hat\Gamma^\alpha_{bc}\CK^b\CK^c-
\overline{\hat\Gamma^\alpha_{bc}\CK^b\CK^c}
=\AA^{,\alpha}-\BB^{\alpha\prime}
- \left( \BB_\beta^{\;\;|\alpha} - \BB^\alpha_{\;\;|\beta}
+ 2 \CC^{\alpha\prime}_\beta \right) e^\beta+ 
\left( 2 \CC^\alpha_{\beta|\gamma} - \CC_{\beta\gamma}^{\;\;\;\;|\alpha}
\right) e^\beta e^\gamma+\hat{\bar\Gamma}^\alpha_{\beta\gamma}
\left(2e^\beta+\dea^\beta\right)\dea^\gamma
\nonumber \\
&&
+2\dnu\left(\AA^{,\alpha} - \BB^{\alpha\prime} \right)
-\left( \dea^\beta + \dnu e^\beta \right)\left(\BB_\beta^{\;\;|\alpha} 
- \BB^\alpha_{\;\;|\beta}+ 2 \CC^{\alpha\prime}_\beta \right)
+ 2 \left( 2 \CC^\alpha_{\beta|\gamma}-\CC_{\beta\gamma}^{\;\;\;\;|\alpha} 
\right)e^\beta \dea^\gamma + \AA' \BB^\alpha- 2 \AA_{,\beta} \CC^{\alpha\beta}
+ 2 \CC^\alpha_\beta \BB^{\beta\prime}\nonumber \\
&&
- 2 \BB^\alpha \AA_{,\beta} e^\beta
+ 4 \CC^{\alpha\gamma} \left( \BB_{[\beta|\gamma]}
+ \CC_{\beta\gamma}' \right) e^\beta
- \left[2~ \CC^\alpha_\delta \left( 2 \CC^\delta_{\beta|\gamma}
- \CC_{\beta\gamma}^{\;\;\;\; |\delta} \right) 
- \BB^\alpha \left( \BB_{\beta|\gamma} + \CC'_{\beta\gamma}
\right)\right] e^\beta e^\gamma  ~. 
\enar
The spatial component of the geodesic equation indicates that the background
photon propagation direction $e^\alpha$ is constant. Another useful quantity is
\beeq
\label{eq:cnu}
\CK^a\CU_a=-\left[1+\dnu+\AA+\left(\VV_\alpha-\BB_\alpha\right)e^\alpha
-{1\over 2}\AA^2+{1\over 2}\VV^\alpha \VV_\alpha+\dnu\AA
+\left(\AA \BB_\alpha+2\VV^\beta \CC_{\alpha\beta}\right)
e^\alpha+(\VV_\alpha-\BB_\alpha)\dea^\alpha
\right]~,
\eneq
and for later reference we define the above quantity as
$\CK^a\CU_a\equiv-(1+\dhnu)$.

\subsection{Observed redshift}
\label{ssec:obsz}
The zero-th order
photon path can be obtained by integrating the photon wave vector
as a function of the affine parameter~$\cc$ as
\beeq
\label{eq:srchomo}
\bar x^a(\cc_s)-\bar x^a(\cc_o)
=\left[\bar\tau_s-\bar\tau_o,~\bar x^\alpha_s\right]
=\left[\cc_s-\cc_o,~(\cc_o-\cc_s)e^\alpha\right]~,
\eneq
where we set $\bar x^\alpha(\cc_o)=\bar x^\alpha_o=0$ and  let
$\bar x^a_s=\bar x^a(\cc_s)$.
Without loss of generality ($\cc\rightarrow \cc+$constant),
we set $\cc_o=0$ hereafter. Therefore, we have the defining
relation between the affine parameter and the line-of-sight distance
\beeq
\cc=\bar\tau-\bar\tau_o=-\rbar=-\int_0^{\zz}{dz'\over H(z')}~,\qquad
\bar\tau_o=\int_0^\infty{dz\over H(z)}~,
\label{eq:affine}
\eneq
where $H(z)=\dot a/a$ is the Hubble parameter and $\zz$ is the redshift
parameter corresponding to the conformal time $\bar\tau$.
Given a redshift parameter~$\zz$, we denote the affine parameter~$\cc_z$,
satisfying
\beeq
1+\zz={1\over a(\bar\tau_z)}~,\qquad\cc_z=\bar\tau_\zz-\bar\tau_o~.
\label{eq:affine2}
\eneq

In an inhomogeneous universe, the positions $x^a_\cc=x^a(\cc)$
of the photon source and the
observer given the affine parameters (or the redshift parameter)
deviate from the positions $\bar x^a_\cc=\bar x^a(\cc)$ 
in a homogeneous universe:
\beeq
x^\alpha_\cc=\bar x^\alpha_\cc+\delta x^\alpha_\cc~,\qquad
\tau_\cc=\bar\tau_\cc+\dT_\cc~.
\eneq
Photons emitted from $x^a_s$ are received by the observer at $x^a_o$,
and the observed redshift~$\zz$ is the ratio of the photon wavelengths
at source and observer
\beeq
\label{eq:obsz}
1+\zz={\left(k^au_a\right)_s\over\left(k^au_a\right)_o}
\equiv{1+\dz\over a(\tau_s)} ~,
\eneq
where we defined the distortion $\dz$ in the observed redshift.
Compared to Eq.~(\ref{eq:affine2}) in a homogeneous universe, 
the observed redshift in 
Eq.~(\ref{eq:obsz}) is affected not only by the expansion of the Universe,
but also by the perturbations such as peculiar velocities of the source and
the observer.  Using Eqs.~(\ref{eq:conf}) and~(\ref{eq:wave}), we derive
\bear
\label{eq:dz}
\dz&=&\DA_o+
\Bigg[\dnu+ \AA +\left( \VV_\alpha - \BB_\alpha\right)
e^\alpha+ \dnu\AA + \dea^\alpha \left(\VV_\alpha - \BB_\alpha \right)
+ \left( \AA \BB_\alpha + 2 \CC_{\alpha\beta} \VV^\beta \right) e^\alpha
+ {1 \over 2} \VV^\alpha\VV_\alpha - {1 \over 2} \AA^2 \Bigg]^s_o \nonumber\\
&&+
\bigg[\DA-\dnu- \AA -\left( \VV_\alpha - \BB_\alpha\right)e^\alpha\bigg]_o
\Bigg[\dnu+ \AA +\left( \VV_\alpha - \BB_\alpha\right)e^\alpha\Bigg]^s_o ~,
\enar
where the brackets with superscript~$s$ and subscript~$o$
represents a difference of the quantities at the source
position $x^a(\cc_s)$ and the observer position $x^a(0)$ and the bracket
with only subscript~$o$ represents the quantity is evaluated at the observer
position. In deriving Eq.~(\ref{eq:dz}) we account for the fact 
that the observer position deviates from that in a homogeneous universe
\beeq
a(\tau_o)=a[\bar\tau(0)+\dT(0)]=1+\HH_o\dT_o+{1\over2}\left(
\HH^2_o+\HH'_o\right)\dT_o^2\equiv1+\DA_o~,
\eneq
where the conformal Hubble parameter is $\HH=aH$,
while the spatial position can be always set zero 
$x^\alpha_o=\bar x^\alpha_o=0$ due to symmetry in a homogeneous universe.

In Eq.~(\ref{eq:dz}), perturbation quantities are evaluated at the source
position $x^a_s$, which is close but not exactly at 
the observed redshift~$\zz$, i.e., $\bar x^a(\cc_z)$.
To facilitate further calculations, we define a perturbation $\Dcc_s$ in the
affine parameter~$\cc_s$ as
$\cc_s\equiv\cc_z+\Dcc_s$, where $\cc_z$ satisfies the
relation in Eq.~(\ref{eq:affine2}). 
To the second order in perturbations, the source position can be rephrased as
\bear
\label{eq:DT}
\tau_s&=&\tau(\cc_z+\Dcc_s)=\bar\tau(\cc_z+\Dcc_s)+\dT(\cc_z+\Dcc_s)=
\bar\tau_z+\Dcc_s+\dT_z+\dnu_z\Dcc_s\equiv\bar\tau_z+\DT_z~,\\ 
\label{eq:DX}
x^\alpha_s&=&x^\alpha(\cc_z+\Dcc_s)=\bar x^\alpha(\cc_z+\Dcc_s)+\delta 
x^\alpha(\cc_z+\Dcc_s)=\bar x^\alpha_z-e^\alpha\Dcc_s
+\delta x^\alpha_z-{1\over2}
\bar\Gamma^\alpha_{\beta\gamma}e^\beta e^\gamma\Dcc_s^2
-\dea^\alpha_z\Dcc_s \nonumber\\
&\equiv&\bar x^\alpha_z+\DX^\alpha_z~,
\enar
where the subscript~$\zz$ indicates that quantities are evaluated at the affine
parameter~$\cc_z$ and $\bar\Gamma^\alpha_{\beta\gamma}$ is the Christoffel
symbol based on $\gbar_{\alpha\beta}$. For the deviation of the source 
position in an inhomogeneous universe, we will need only the first order
terms in $\DX^a_z\equiv(\DT_z,\DX_z^\alpha)$, and we will compute it 
in detail in Sec.~\ref{ssec:path}. The distortion in the observed redshift is,
therefore, 
\bear
\label{eq:dzone}
\dz^{(1)}&=&\HH_o\dT_o+
\Bigg[\dnu+ \AA +\left( \VV_\alpha - \BB_\alpha\right)e^\alpha\Bigg]^z_o ~,\\
\label{eq:dztwo}
\dz^{(2)}&=&\DA_o+\Bigg[\dnu+ \AA +\left( \VV_\alpha - \BB_\alpha\right)
e^\alpha+\dnu\AA + \dea^\alpha \left(\VV_\alpha - \BB_\alpha \right)
+ \left( \AA \BB_\alpha + 2 \CC_{\alpha\beta} \VV^\beta \right) e^\alpha
+ {1 \over 2} \VV^\alpha\VV_\alpha - {1 \over 2} \AA^2 \Bigg]^z_o \nonumber\\
&&
+\bigg[\DA-\dnu- \AA -\left( \VV_\alpha - \BB_\alpha\right)e^\alpha\bigg]_o
\Bigg[\dnu+ \AA +\left( \VV_\alpha - \BB_\alpha\right)e^\alpha\Bigg]^z_o 
+\DX_z^b\bigg[\dnu+ \AA +\left( \VV_\alpha - \BB_\alpha\right)e^\alpha
\bigg]_{,b}~,
\enar
where we omitted the superscripts of the perturbation orders for simplicity.
Consistently to the second order in perturbations,
first-order and second-order perturbation quantities in Eqs.~(\ref{eq:dzone})
and~(\ref{eq:dztwo}) at the source position can be evaluated at the 
observed redshift, while the
former results in additional second-order contributions due to the first-order
deviation of the source position from the observed redshift.
Similar calculations can be found in \cite{PYCA96,BEMAET12a}.

So far, we left unspecified the perturbation $\Dcc_s$ in the affine parameter.
Using the defining relation in Eq.~(\ref{eq:affine2}), the observed redshift
in Eq.~(\ref{eq:obsz}) can be written as
\beeq
1+\zz={1\over a(\bar\tau_\zz)}={1+\dz\over a\left[\tau(\cc_z+\Dcc_s)\right]}~,
\label{eq:matchz}
\eneq
and note that the observed redshift is independent of how we label the
source position using the affine parameter. 
Substituting Eq.~(\ref{eq:DT}) into Eq.~(\ref{eq:matchz}) yields that
the perturbation $\Dcc_s$ in the affine parameter satisfies
\beeq
\HH_z\DT_z+{1\over2}(\HH_z^2+\HH'_z)(\Dcc_s+\dT_z)^2=\dz~,
\eneq
and we derive
\beeq
\Dcc_s^{(1)}=-\dT_z^{(1)}+{\dz^{(1)}\over\HH_z}~,\quad
\Dcc_s^{(2)}=-\dT_z^{(2)}-\dnu_z\left(-\dT_z+{\dz\over\HH_z}
\right)-{1\over2\HH_z^3}\left(\HH_z^2+\HH_z'\right)\dz^2
+{\dz^{(2)}\over\HH_z}~,
\eneq
where the perturbation quantities in quadratic form are evaluated at the
linear order and $\HH_z=\HH(z)$.

\subsection{Observed angle of source galaxies}
\label{ssec:angle}
The observed source position in the sky is described by the observed angle
$\nhat=(\ttt,\pp)$ in the local observer frame. In a homogeneous universe, 
it is identical to the unit directional vector $e^\alpha$. However, 
the observer frame is moving in an inhomogeneous universe, and these two
unit directional vectors are different, simply because of the change of frame.
Therefore, it is necessary to express the source galaxy position, not
only in terms of the observed redshift~$\zz$, but also in terms of the
observed angle $(\ttt,\pp)$. 

In Appendix~\ref{app:angle}, we explicitly derive the photon wavevector~$k^a$
in the FRW frame by transforming the observed photon wavevector
in the observer's rest frame. The photon wavevector is completely
set by local observables quantities such as the photon frequency~$\nu$ 
and the angle $(\ttt,\pp)$.
However, with additional degree of freedom $\NC$ in the
conformally transformed wavevector in Eq.~(\ref{eq:conf}), we can choose
the normalization of the photon wavevector~$\CK^a$ to
 simplify the calculations by aligning the two unit
directional vectors $n^\alpha=e^\alpha$. Though the choice has no
impact on the description of observable quantities, other choice would
make the calculation significantly complicated. The detailed calculations
are presented in Appendix~\ref{app:photon}.

\subsection{Distortions in photon path}
\label{ssec:path}
Having computed the observed redshift and the observed angle, we 
now express the source position in terms of metric 
perturbations. In the presence of perturbations in an inhomogeneous universe, 
the photon path at the affine parameter~$\cc$ is distorted as
\beeq
x^a_\cc-x^a_o=\left[\tau_\cc-\tau_o~,~x^\alpha_\cc\right]
=\left[\cc+\int_0^\cc d\cc'~\dnu~,~-\cc e^\alpha-\int_0^\cc d\cc'~
\dea^\alpha\right]~,
\label{eq:xpert}
\eneq
and the deviation of the position from that in a homogeneous universe is
\beeq
\delta x^a=x^a_\cc-\bar x^a_\cc=\left[\dT_\cc~,~\delta x^\alpha_\cc\right]
=\left[\dT_o~,~0\right]+\left[\int_0^\cc d\cc'~\dnu~,~-\int_0^\cc d\cc'~
\dea^\alpha\right]~,
\label{eq:dx}
\eneq
where the integration over the affine parameter ($d\cc$)
represents the integration
along the photon path $x^a(\cc)$, not necessarily along the straight line
$\bar x^a(\cc)$. Note that
the affine parameter is defined as a parameter without resort to
whether we consider homogeneous or inhomogeneous universes. Using the
geodesic equations in~(\ref{eq:tempo}) and~(\ref{eq:spatial}), we derive
the perturbations in the photon wavevector as
\bear
\label{eq:formal}
\dnu_\cc-\dnu_o&=&-\int_0^\cc d\cc'~\GG^0~,\qquad
\dea^\alpha_\cc-\dea^\alpha_o=\int_0^\cc d\cc'~\GG^\alpha~,\\
\dnu^{(1)}\bigg|_o^\zz&=&
-2(\AA_z-\AA_o)-\int_0^{\rbar_z} d\rbar~\left[\AA'-\left(\BB_{\alpha|\beta}
+\CC_{\alpha\beta}'\right)e^\alpha e^\beta\right]~,\\
\dea^{\alpha(1)}\bigg|_o^\zz&=&
-\bigg[\BB^\alpha+2\CC^\alpha_\beta e^\beta\bigg]^\zz_o-\int_0^{\rbar_z}
d\rbar~\left(\AA-\BB_\beta e^\beta-\CC_{\beta\gamma}e^\beta e^\gamma\right)
^{|\alpha}~,\\
\dnu^{(2)}\bigg|_o^\zz&=&-2\left(\AA_z-\AA_o\right)
-\int_0^{\rbar_z}d\rbar~\bigg\{
\AA'-\left(\BB_{\alpha|\beta}+\CC_{\alpha\beta}'\right) e^\alpha e^\beta 
+2\dnu\AA_{,\alpha}e^\alpha +2\AA\AA' \\
&&
+\AA_{,\alpha}\BB^\alpha-\BB^\alpha\BB_\alpha' 
-2\left(2\AA\AA_{,\alpha}+\BB_\beta\CC_\alpha^{\beta\prime}
+\BB^\beta \BB_{[\alpha|\beta]} \right) e^\alpha 
-2~\dea^\alpha \left(\BB_{(\alpha|\beta)} +\CC_{\alpha\beta}'\right)e^\beta
\nonumber \\
&&
+ \left[\AA\left( 2\BB_{\alpha|\beta} 
+2\CC_{\alpha\beta}'\right)+\BB_\gamma \left( 2 \CC^\gamma_{\alpha|\beta}
-\CC_{\alpha\beta}^{\;\;\;\;|\gamma}\right)\right]e^\alpha e^\beta
+\DX^c\left[
\AA'-\left(\BB_{\alpha|\beta}+\CC_{\alpha\beta}'\right) e^\alpha e^\beta 
\right]_{,c}\bigg\}~, \nonumber\\
\dea^{\alpha(2)}\bigg|_o^\zz&=&
-\bigg[\BB^\alpha+2\CC^\alpha_\beta e^\beta\bigg]^\zz_o-\int_0^{\rbar_z}
d\rbar\bigg\{
\left(\AA- \BB_\beta e^\beta- \CC_{\beta\gamma}e^\beta e^\gamma
\right)^{|\alpha}+\dnu\left(2\AA^{,\alpha} - \BB^{\alpha\prime} \right)
-\BB_\beta^{\;\;|\alpha}\left(\dea^\beta + \dnu ~e^\beta\right)
\nonumber \\
&&
+ \dnu\BB^\alpha_{\;\;|\beta}e^\beta- 2 \CC^{\alpha\prime}_\beta\dea^\beta
+ 2\left( \CC^\alpha_{\beta|\gamma}-\CC_{\beta\gamma}^{\;\;\;\;|\alpha} 
\right)e^\beta \dea^\gamma + \AA' \BB^\alpha- 2 \AA_{,\beta} \CC^{\alpha\beta}
+ 2 \CC^\alpha_\beta \BB^{\beta\prime}\nonumber \\
&&
- 2 \BB^\alpha \AA_{,\beta} e^\beta
+ 4 \CC^{\alpha\gamma} \left( \BB_{[\beta|\gamma]}
+ \CC_{\beta\gamma}' \right) e^\beta
- \left[2~ \CC^\alpha_\delta \left( 2 \CC^\delta_{\beta|\gamma}
- \CC_{\beta\gamma}^{\;\;\;\; |\delta} \right) 
- \BB^\alpha \left( \BB_{\beta|\gamma} + \CC'_{\beta\gamma}
\right)\right] e^\beta e^\gamma  \nonumber\\
&&+\DX^d\left(\AA^{,\alpha}- \BB_\beta^{\;\;|\alpha} e^\beta
- \CC_{\beta\gamma}^{\;\;\;\;|\alpha}e^\beta e^\gamma\right)_{,d}\bigg\}~,
\enar
where we used the total derivative with respect to the affine parameter
along the photon path in 
Eq.~(\ref{eq:dlambda}) for simplification. 
The photon path in Eq.~(\ref{eq:xpert}) can be further related to
the integration over the metric perturbations defined in Eqs.~(\ref{eq:tempo})
and~(\ref{eq:spatial}) as
\beeq
x^a_\cc-x^a_o=\left[(1+÷÷÷\dnu_o)\cc-\int_0^\cc d\cc'(\cc-\cc')~\GG^0~,
~-\cc (e^\alpha+\dea^\alpha_o)-\int_0^\cc d\cc'(\cc-\cc')~\GG^\alpha\right]~.
\label{eq:xpert2}
\eneq
Noting that the source position is parametrized by $\cc_s=\cc_z+\Dcc_s$,
we have the source position
\bear
\label{eq:srczero}
\bar x^a_s&=&[\cc_z+\bar\tau_o~,~-\cc_ze^\alpha]
=[\bar\tau_z~,~\rbar_ze^\alpha]~,\\
\label{eq:srcone}
x^{a(1)}_s&=&\left[\dT_o+\Dcc_s-\int_0^{\rbar_z}d\rbar~\dnu~,~-\Dcc_s e^\alpha
+\int_0^{\rbar_z}d\rbar~\dea^\alpha\right]\nonumber \\
&=&\left[\dT_o+\cc_z\dnu_o+\Dcc_s-\int_0^{\rbar_z}d\rbar~(\rbar_z-\rbar)
~\GG^0~,~\rbar_z\dea^\alpha_o-\Dcc_s e^\alpha-\int_0^{\rbar_z}d\rbar~
(\rbar_z-\rbar)~\GG^\alpha\right]~,\\
\label{eq:srctwo}
\tau^{(2)}_s&=&\dT_o+\Dcc_s(1+\dnu_z)-\int_0^{\rbar_z}d\rbar\left(\dnu
+\DX^a~\dnu_{,a}\right)\nonumber \\
&=&\dT_o+\cc_z~\dnu_o+\Dcc_s(1+\dnu_z)
-\int_0^{\rbar_z}d\rbar~\Bigg[(\rbar_z-\rbar)\left(\GG^0+\DX^a~\GG^0_{\;\;,a}
\right)\Bigg]~,\\
\label{eq:srctwo2}
x^{\alpha(2)}_s&=&-\Dcc_s(e^\alpha+\dea^\alpha_z)+\int_0^{\rbar_z}d\rbar
\left(\dea^\alpha+\DX^b\dea^\alpha_{\;\;,b}\right)\nonumber \\
&=&\rbar_z~\dea^\alpha_o-\Dcc_s (e^\alpha+\dea^\alpha_z)
-\int_0^{\rbar_z}d\rbar~\Bigg[(\rbar_z-\rbar)\left(\GG^\alpha+\DX^b~
\GG^\alpha_{\;\;,b}\right)\Bigg]~,
\enar
where the line-of-sight integration here represents
the integration over the unperturbed photon path~$d\rbar$. 

Since the observers identify the source position by measuring the observed 
redshift~$\zz$ and the observed angle $(\ttt,\pp)$, the inferred source 
position is in rectangular coordinates
\beeq
\label{eq:inferred}
\hat x^a_s=[\bar\tau_z~,~\rbar_z \nhat]
=[\bar\tau_z~,~\rbar_z\sin\ttt\cos\pp~,~\rbar_z\sin\ttt\sin\pp~,
~\rbar_z\cos\ttt]~,
\eneq
where $\rbar_z=\rbar(\zz)=\bar\tau_o-\bar\tau_z$, 
$\nhat=(\sin\ttt\cos\pp~,\sin\ttt\sin\pp~,\cos\ttt)$ is a unit 
directional vector based on the observed angle $(\ttt,\pp)$ of the source
and $\bar\tau_z$ is the conformal time defined in Eq.~(\ref{eq:affine2}).
Note that in general $\bar x^a_z\neq\hat x^a_s$ because of the difference
between $e^\alpha$ and $n^\alpha$, but with our choice of normalization
constant we have $\bar x^a_z=\hat x^a_s$.
Given the source position in Eqs.~(\ref{eq:srczero})$-$(\ref{eq:srctwo2}),
we define the distortion $(\dTau,\drr,\dtt,\dpp)$
of the source position $x^a_s$ with respect to the
inferred source position $\hat x^a_s$ by
\beeq
\label{eq:displ}
x^a_s\equiv\left[\bar\tau_z+\dTau
~,~(\rbar_z+\drr)\sin(\ttt+\dtt)\cos(\pp+\dpp)~,~(\rbar_z+\drr)
\sin(\ttt+\dtt)\sin(\pp+\dpp)~,~(\rbar_z+\drr)\cos(\ttt+\dtt)\right]~,
\eneq
where  the deviation in the conformal time of the source position
$\dTau\equiv\tau_s-\bar\tau_z=\DT_z$ is different from $\drr$.\footnote{While
the source position is on the past light cone, the separation of coordinates
and metric perturbations is arbitrary and gauge-dependent. Furthermore,
$\drr$ only represents the radial displacement.}
Since the source position~$x^a_s$ is unobservable, these deviations 
from the inferred position $\hat x^a_s$ are gauge-dependent \cite{YOO10}.
While Eqs.~(\ref{eq:srczero})$-$(\ref{eq:srctwo2}) are valid in general
coordinates, it is most convenient to evaluate the distortions
in rectangular coordinates. 

Constructing two additional unit directional vectors
$\thatv=(\cos\ttt\cos\pp~,\cos\ttt\sin\pp~,-\sin\ttt)$ and
$\phatv=(-\sin\pp~,\cos\pp~,0)$ based on the observed angle,
the distortions of the source position in spherical coordinates are
to the linear order in perturbations  
\bear
\drr^{(1)}&=&(n_\alpha x^\alpha_s)^{(1)}=
-\Dcc_s+\int_0^{\rbar_z}d\rbar~e_\alpha\dea^\alpha=
\dT_o-{\dz\over\HH_z}+\int_0^{\rbar_z}d\rbar~\left(
\AA-\BB_\alpha e^\alpha-\CC_{\alpha\beta}e^\alpha e^\beta \right)~,\\
\rbar_z\dtt^{(1)}&=&(\ttt_\alpha x^\alpha_s)^{(1)}=
\rbar_ze_{\theta\alpha}\dea^\alpha_o
-\int_0^{\rbar_z}d\rbar~(\rbar_z-\rbar)~e_{\theta\alpha}\GG^\alpha\\
&=&\rbar_ze_{\theta\alpha}\left(\dea^\alpha+\BB^\alpha+2~\CC^\alpha_\beta
e^\beta\right)_o-\int_0^{\rbar_z}d\rbar~\left[e_{\theta\alpha}
\left(\BB^\alpha+2~\CC^\alpha_\beta e^\beta\right)+\left({\rbar_z-\rbar\over
\rbar}\right){\partial\over\partial\ttt}
\left(\AA-\BB_\alpha e^\alpha-\CC_{\alpha\beta}e^\alpha e^\beta \right)
\right]~,\nonumber\\
\rbar_z\sin\ttt~\dpp^{(1)}&=&(\pp_\alpha x^\alpha_s)^{(1)}=
\rbar_ze_{\phi\alpha}\dea^\alpha_o
-\int_0^{\rbar_z}d\rbar~(\rbar_z-\rbar)~e_{\phi\alpha}\GG^\alpha\\
&=&\rbar_ze_{\pp\alpha}\left(\dea^\alpha+\BB^\alpha+2~\CC^\alpha_\beta
e^\beta\right)_o-\int_0^{\rbar_z}d\rbar~\left[e_{\pp\alpha}
\left(\BB^\alpha+2~\CC^\alpha_\beta e^\beta\right)+\left({\rbar_z-\rbar\over
\rbar\sin\ttt}\right){\partial\over\partial\pp}
\left(\AA-\BB_\alpha e^\alpha-\CC_{\alpha\beta}e^\alpha e^\beta \right)
\right]~,\nonumber
\enar
and to the second order in perturbations
\bear
\drr^{(2)}&=&(n_\alpha x^\alpha_s)^{(2)}+{1\over2\rbar_z}\left[
(\ttt_\alpha x^\alpha_s)^2+(\pp_\alpha x^\alpha_s)^2\right]~,\\
\rbar_z~\dtt^{(2)}&=&(\ttt_\alpha x^\alpha_s)^{(2)}
-{(n_\alpha x^\alpha_s)
(\ttt_\alpha x^\alpha_s)\over\rbar_z}
+{\cot\ttt\over2\rbar_z}(\pp_\alpha x^\alpha_s)^2~, \\
\rbar_z\sin\ttt~\dpp^{(2)}&=&(\pp_\alpha x^\alpha_s)^{(2)}
-{(n_\alpha x^\alpha_s)(\pp_\alpha x^\alpha_s)\over\rbar_z}
-{\cot\ttt(\ttt_\alpha x^\alpha_s)(\pp_\alpha x^\alpha_s)\over \rbar_z}~,
\enar
where the quadratic terms are at the first order and the remaining second-order
pieces are
\bear
(n_\alpha x^\alpha_s)^{(2)}&=&-\Dcc_s^{(2)} -\Dcc_se_\alpha\dea^\alpha_z
+\int_0^{\rbar_z}d\rbar~\Bigg[e_\alpha\dea^{\alpha(2)}
+e_\alpha\DX^b\dea^\alpha_{\;\;,b}\Bigg]\\
&=&\dT_o^{(2)}+\left(\dT_z-{\dz\over\HH_z}\right)\left(\AA-\BB_\alpha e^\alpha-
\CC_{\alpha\beta}e^\alpha e^\beta\right)_\zz
+{1\over2\HH_z^3}\left(\HH_z^2+\HH_z'\right)\dz^2-{\dz^{(2)}\over\HH_z}
\nonumber \\
&&+\int_0^{\rbar_z}d\rbar~\bigg\{
\left(\AA-\BB_\alpha e^\alpha-\CC_{\alpha\beta}~e^\alpha e^\beta\right)^{(2)}
+\dnu\bigg[{1\over2}\dnu+\left(2\AA-\BB_\alpha e^\alpha\right)\bigg]
-\bigg[{1\over2}\dea_\alpha
+\left(\BB_\alpha+2~\CC_{\alpha\beta}~e^\beta\right)\bigg]\dea^\alpha
\nonumber \\
&&\hspace{50pt}
+\DX^c\left(\AA-\BB_\alpha e^\alpha-\CC_{\alpha\beta}~e^\alpha e^\beta
\right)_{,c}\bigg\}
~,\nonumber \\
(\ttt_\alpha x^\alpha_s)^{(2)}&=&\rbar_z~e_{\theta\alpha}\dea^{\alpha(2)}_o
-\Dcc_s e_{\theta\alpha}\dea^\alpha_z
-\int_0^{\rbar_z}d\rbar~
\Bigg[(\rbar_z-\rbar)e_{\theta\alpha}\left(\GG^\alpha+\DX^b~
\GG^\alpha_{\;\;,b}\right)\Bigg]\\
&=&\rbar_z e_{\ttt\alpha}\left(\dea^\alpha+\BB^\alpha
+2~\CC^\alpha_\beta e^\beta\right)_o^{(2)}-\Dcc_s e_{\ttt\alpha}\dea^\alpha_z
-\int_0^{\rbar_z}d\rbar~\bigg\{e_{\ttt\alpha}\left(1+\DX^b\partial_b\right)
\left(\BB^\alpha+2~\CC^\alpha_\beta e^\beta\right)
\nonumber \\
&&+\left({\rbar_z-\rbar\over\rbar}\right)
{\partial\over\partial\ttt}\bigg[(1+\DX^b\partial_b)
\left(\AA-\BB_\alpha e^\alpha-\CC_{\alpha\beta}e^\alpha e^\beta\right) \bigg]
+(\rbar_z-\rbar)e_{\ttt\alpha}\bigg[
\dnu\left(2\AA^{,\alpha} - \BB^{\alpha\prime} \right)
-\BB_\beta^{\;\;|\alpha}\left(\dea^\beta + \dnu ~e^\beta\right)
\nonumber \\
&&
+ \dnu\BB^\alpha_{\;\;|\beta}e^\beta- 2 \CC^{\alpha\prime}_\beta\dea^\beta
+ 2\left( \CC^\alpha_{\beta|\gamma}-\CC_{\beta\gamma}^{\;\;\;\;|\alpha} 
\right)e^\beta \dea^\gamma + \AA' \BB^\alpha- 2 \AA_{,\beta} \CC^{\alpha\beta}
+ 2 \CC^\alpha_\beta \BB^{\beta\prime}- 2 \BB^\alpha \AA_{,\beta} e^\beta
\nonumber\\
&&
+ 4 \CC^{\alpha\gamma} \left( \BB_{[\beta|\gamma]}
+ \CC_{\beta\gamma}' \right) e^\beta
- \left[2~ \CC^\alpha_\delta \left( 2 \CC^\delta_{\beta|\gamma}
- \CC_{\beta\gamma}^{\;\;\;\; |\delta} \right) 
- \BB^\alpha \left( \BB_{\beta|\gamma} + \CC'_{\beta\gamma}
\right)\right] e^\beta e^\gamma \bigg]\bigg\}~,\nonumber \\
(\pp_\alpha x^\alpha_s)^{(2)}&=&
\rbar_z~e_{\phi\alpha}\dea^{\alpha(2)}_o-\Dcc_s e_{\phi\alpha}\dea^\alpha_z
-\int_0^{\rbar_z}d\rbar~\Bigg[(\rbar_z-\rbar)
e_{\phi\alpha}\left(\GG^\alpha+\DX^b~
\GG^\alpha_{\;\;,b}\right)\Bigg]\\
&=&\rbar_z e_{\pp\alpha}\left(\dea^\alpha+\BB^\alpha
+2~\CC^\alpha_\beta e^\beta\right)_o^{(2)}-\Dcc_s e_{\pp\alpha}\dea^\alpha_z
-\int_0^{\rbar_z}d\rbar~\bigg\{e_{\pp\alpha}\left(1+\DX^b\partial_b\right)
\left(\BB^\alpha+2~\CC^\alpha_\beta e^\beta\right)
\nonumber \\
&&+\left({\rbar_z-\rbar\over\rbar\sin\ttt}\right)
{\partial\over\partial\pp}\bigg[(1+\DX^b\partial_b)
\left(\AA-\BB_\alpha e^\alpha-\CC_{\alpha\beta}e^\alpha e^\beta\right) \bigg]
+(\rbar_z-\rbar)e_{\pp\alpha}\bigg[
\dnu\left(2\AA^{,\alpha} - \BB^{\alpha\prime} \right)
-\BB_\beta^{\;\;|\alpha}\left(\dea^\beta + \dnu ~e^\beta\right)
\nonumber \\
&&
+ \dnu\BB^\alpha_{\;\;|\beta}e^\beta- 2 \CC^{\alpha\prime}_\beta\dea^\beta
+ 2\left( \CC^\alpha_{\beta|\gamma}-\CC_{\beta\gamma}^{\;\;\;\;|\alpha} 
\right)e^\beta \dea^\gamma + \AA' \BB^\alpha- 2 \AA_{,\beta} \CC^{\alpha\beta}
+ 2 \CC^\alpha_\beta \BB^{\beta\prime}- 2 \BB^\alpha \AA_{,\beta} e^\beta
\nonumber\\
&&
+ 4 \CC^{\alpha\gamma} \left( \BB_{[\beta|\gamma]}
+ \CC_{\beta\gamma}' \right) e^\beta
- \left[2~ \CC^\alpha_\delta \left( 2 \CC^\delta_{\beta|\gamma}
- \CC_{\beta\gamma}^{\;\;\;\; |\delta} \right) 
- \BB^\alpha \left( \BB_{\beta|\gamma} + \CC'_{\beta\gamma}
\right)\right] e^\beta e^\gamma \bigg]\bigg\}~.\nonumber 
\enar
The distortions of the source position are decomposed as the radial and
the angular displacements. Both of them arise due to the metric perturbations
along the photon path, and the identification of the source at the observed 
redshift contributes to the radial displacement.

\subsection{Lensing magnification}
\label{ssec:kappa}
The distortion in the solid angle $d\Omega$ at the observed $(\ttt,\pp)$
and the (unobserved) source $(\ttt+\dtt,\pp+\dpp)$ is described by the
deformation matrix~$\DD$ (inverse of the magnification matrix), and it is
conventionally decomposed as
\beeq
\label{eq:deform}
\DD={\partial(\ttt+\dtt,\pp+\dpp)\over \partial(\ttt,\pp)}\equiv
I-\left(\begin{array}{cc}\kappa&0\\0&\kappa\end{array}\right)
-\left(\begin{array}{cc}0&\omega\\-\omega&0\end{array}\right)
-\left(\begin{array}{cc}\gamma_1&\gamma_2\\\gamma_2&-\gamma_1\end{array}\right)
~,
\eneq
where $\kappa$ is the gravitational lensing convergence, $\omega$ is the
rotation, and $(\gamma_1,\gamma_2)$ is the shear (e.g., see 
\cite{SCEHFA92,BASC01,MUVAET08} for reviews).
The ratio of the solid angles is the Jacobian of the angular transformation 
or the determinant of the deformation matrix:
\beeq
\label{eq:Ddet}
\up{det}~\DD=1-2~\kappa+\kappa^2-\gamma^2+\omega^2\equiv1-2~\kappa^{(1)}
+\dDD^{(2)}~,
\eneq
where we defined the second-order part $\dDD$ of the determinant. Note that the
first-order term is simply the gravitational lensing convergence, and
we only need the determinant term,
not the individual components of shear and rotation.
To the second order in perturbations, the determinant of the deformation
matrix is
\beeq
\label{eq:kappadef}
\up{det}~\DD
={\sin(\ttt+\dtt)\over \sin\ttt}
\left[1+{\partial\over\partial\ttt}\dtt+{\partial\over\partial\pp}\dpp
+{\partial\over\partial\ttt}\dtt{\partial\over\partial\pp}\dpp
-{\partial\over\partial\ttt}\dpp{\partial\over\partial\pp}\dtt
\right]~,
\eneq
yielding the relation 
\bear
\kappa^{(1)}&=&-{1\over2}\left[\left(\cot\ttt+{\partial\over\partial\ttt}
\right)\dtt+{\partial\over\partial\pp}\dpp\right]~,\\
\dDD^{(2)}&=&\left(\cot\ttt+{\partial\over\partial\ttt}
\right)\dtt+{\partial\over\partial\pp}\dpp
+{\partial\over\partial\ttt}\dtt{\partial\over\partial\pp}\dpp
-{\partial\over\partial\ttt}\dpp{\partial\over\partial\pp}\dtt-{1\over2}\dtt^2
+\cot\ttt~\dtt\left({\partial\over\partial\ttt}\dtt
+{\partial\over\partial\pp}\dpp\right)~.
\enar
The second-order calculations of the gravitational lensing and shear
can be found in \citet{BEBOVE10,BEBOET12}. These expressions can be 
further related to the metric perturbations by using the distortions 
in the photon path computed in Sec.~\ref{ssec:path}, but they provide a more
physical transparent intuition as written in terms of the angular 
displacements ($\dtt,\dpp$) of the source galaxy position.

\subsection{Observed volume element}
\label{ssec:dV}
Having computed the distortion in photon path,
we are now in a position to compute the physical volume occupied by the 
observed source galaxies over the small intervals $d\zz$ in observed redshift
and $(d\ttt,d\pp)$ in observed angle, and to express the volume in terms of
the observed quantities. Since the real position $x^a_s$
of source galaxies is parametrized by using the observed quantities,
the physical volume in the rest frame of the observed source galaxies can
be written in a covariant way as \cite{WEINB72,YOFIZA09,YOO10}
\bear
\label{eq:dv}
dV_\up{phy}
&=&\sqrt{-g}~\varepsilon_{dabc}~\UU^d_s~{\partial x^a_s\over\partial\zz}
{\partial x^b_s\over\partial\ttt}{\partial x^c_s\over\partial\pp}~
d\zz~d\ttt~d\pp \\
&=&{(1+\dz)^3\over(1+\zz)^3}(1+\dg)\left[
\varepsilon_{0\alpha\beta\gamma}~{\partial x^\alpha\over\partial\zz}
{\partial x^\beta\over\partial\ttt}{\partial x^\gamma\over\partial\pp}
+\varepsilon_{0\alpha\beta\gamma}~\dUU^0
~{\partial x^\alpha\over\partial\zz}
{\partial x^\beta\over\partial\ttt}{\partial x^\gamma\over\partial\pp}
+\varepsilon_{\delta abc}~\VV^\delta~{\partial x^a\over\partial\zz}
{\partial x^b\over\partial\ttt}{\partial x^c\over\partial\pp}
\right]d\zz~d\ttt~d\pp ~,\nonumber 
\enar
where the subscript~$s$ for the source position is omitted in the second line,
and we simply expanded the summation of the Levi-Civita symbol 
$\varepsilon_{abcd}$ over the four velocity for further calculations.
The distortion $\dz$ in the observed redshift is given
in Eqs.~(\ref{eq:dzone}) and~(\ref{eq:dztwo}), the four velocity $\UU^a$ is
given in Eq.~(\ref{eq:fourv}), and finally the metric determinant is
\beeq
\label{eq:ggg}
\sqrt{-g}\equiv a^4(1+\dg)~,\qquad
\dg=\AA+\CC^\alpha_\alpha-{1\over2}\AA^2+\BB^\alpha\BB_\alpha
+\AA~\CC^\alpha_\alpha+{1\over2}\left(\CC^\alpha_\alpha\right)^2-
\CC^\alpha_\beta\CC^\beta_\alpha~.
\eneq

To the second order in perturbations, we compute the individual
terms in the square
bracket in Eq.~(\ref{eq:dv}). First, the last term in the square bracket is 
\beeq
\varepsilon_{\delta abc}~\VV^\delta~{\partial x^a\over\partial\zz}
{\partial x^b\over\partial\ttt}{\partial x^c\over\partial\pp}=
{\rbar^2_z\sin\ttt\over H_z}\left\{\VP+\VP\left[2~{\drr\over\rbar_z}
-2~\kappa-H_z{\partial\over\partial\zz}~\dTau\right]
-{1\over\rbar_z}\left(\VTT{\partial\over\partial\ttt}
+{\VPP\over\sin\ttt}{\partial\over\partial\pp}\right)
\left(\drr+\dTau\right)+\VTT~\dtt+\VPP\sin\ttt~\dpp\right\}~,
\eneq
and the second term is
\beeq
\varepsilon_{0\alpha\beta\gamma}~\dUU^0
~{\partial x^\alpha\over\partial\zz}
{\partial x^\beta\over\partial\ttt}{\partial x^\gamma\over\partial\pp}=
{\rbar^2_z\sin\ttt\over H_z}\left\{
\dUU^0+\dUU^0\left[2~{\drr\over\rbar_z}-2~\kappa
+H_z{\partial\over\partial\zz}~\drr\right]\right\}~,
\eneq
where the spatial component of the source four velocity is decomposed into
the line-of-sight and the transverse velocities
\beeq
\label{eq:vvdec}
\VV^\alpha\equiv\VP n^\alpha+\VTT\ttt^\alpha+\VPP\pp^\alpha~.
\eneq
These two terms in the square bracket in Eq.~(\ref{eq:dv})
vanish in a homogeneous universe. Finally, 
the first term in the square bracket is
\bear
\varepsilon_{0\alpha\beta\gamma}~{\partial x^\alpha\over\partial\zz}
{\partial x^\beta\over\partial\ttt}{\partial x^\gamma\over\partial\pp}
&=&{\rbar^2_z\sin\ttt\over H_z}\Bigg\{1+2~{\drr\over\rbar_z}-2~\kappa
+H_z{\partial\over\partial\zz}~\drr
+{\drr^2\over\rbar^2_z}+2~{\drr\over\rbar_z}
\left(H_z{\partial\over\partial\zz}~\drr-2~\kappa\right) \\
&&-2H_z\kappa~{\partial\over\partial\zz}\drr
-H_z{\partial\over\partial\zz}~\dtt~{\partial\over\partial\ttt}~\drr
-H_z{\partial\over\partial\zz}~\dpp~{\partial\over\partial\pp}~\drr
\Bigg\}~.\nonumber
\enar
Summing up the individual contributions, we obtain the physical volume 
defined in Eq.~(\ref{eq:dv}) as
\beeq
\label{eq:ddv}
dV_\up{phy}
=\sqrt{-g}~\varepsilon_{dabc}~\UU^d_s~{\partial x^a_s\over\partial\zz}
{\partial x^b_s\over\partial\ttt}{\partial x^c_s\over\partial\pp}~
d\zz~d\ttt~d\pp
\equiv{\rbar_z^2~\sin\ttt\over H_z(1+\zz)^3}~d\zz ~d\ttt ~d\pp~(1+\dV)~,
\eneq
and derive the volume distortion  
\bear
\label{eq:ddv1}
\dV^{(1)}&=&3~\dz+\dg+2~{\drr\over\rbar_z}-2~\kappa
+H_z{\partial\over\partial\zz}~\drr+\dUU^0+\VP~,\\
\label{eq:ddv2}
\dV^{(2)}&=& 3~\dz+\dg+2~{\drr\over\rbar_z}+\dDD
+H_z{\partial\over\partial\zz}~\drr+\dUU^0+\VP
+\dUU^0\left[2~{\drr\over\rbar_z}-2~\kappa
+H_z{\partial\over\partial\zz}~\drr\right]\nonumber \\
&&
+{\drr^2\over\rbar^2_z}+2~{\drr\over\rbar_z}
\left(H_z{\partial\over\partial\zz}~\drr-2~\kappa\right) 
-2H_z\kappa~{\partial\over\partial\zz}\drr
-H_z{\partial\over\partial\zz}~\dtt~{\partial\over\partial\ttt}~\drr
-H_z{\partial\over\partial\zz}~\dpp~{\partial\over\partial\pp}~\drr \nonumber\\
&&
+\VP\left[2~{\drr\over\rbar_z}
-2~\kappa-H_z{\partial\over\partial\zz}~\dTau\right]
-{1\over\rbar_z}\left(\VTT{\partial\over\partial\ttt}
+{\VPP\over\sin\ttt}{\partial\over\partial\pp}\right)
\left(\drr+\dTau\right)+\VTT~\dtt+\VPP\sin\ttt~\dpp \nonumber \\
&&+(\dg+3~\dz)
\left[2~{\drr\over\rbar_z}-2~\kappa+H_z{\partial\over\partial\zz}~\drr
+\dUU^0+\VP\right] +3~\dz~\dg+3~\dz^2
+\DX^b\dV^{(1)}_{,b} ~,
\enar
where the perturbation quantities are now evaluated at the observed redshift
and additional 2nd-order terms are added due to the 1st-order deviation of
the photon path. It is noted that the partial derivatives with respect to
the observed quantities ($\zz,\ttt,\pp$) are the partial derivatives with other
observed quantities fixed; The derivative with respect to the observed
redshift is the line-of-sight derivative along the past light cone, involving
not only the spatial derivative, but also the time derivative, while 
the observed angular position ($\ttt,\pp$) is fixed.

The volume distortion to the linear order has a simple physical interpretation
as the distortion compared to the volume element in a homogeneous universe in 
Eq.~(\ref{eq:ddv}) --- $3~\dz$ from the comoving factor $(1+\zz)^3$,
$\dg$, $\dUU^0$, and $\VP$ from defining the source rest frame,
$2~\drr/\rbar_z$ from the volume factor $\rbar^2_z$, $2~\kappa$ from the
solid angle $d\Omega$, and $H_z\partial_z\drr$ from the change of the
radial displacement at the observed redshift. To the second order in 
perturbations, these physical interpretations remain valid 
in the second-order volume distortion. However, additional physical effects
need to be taken into account such as the contribution
of the source tangential velocity and the tangential variation of the
source position, similar to the transverse Doppler effect. 
There exist, of course, nonlinear coupling terms 
with the linear-order volume distortion.

\subsection{Fluctuation in luminosity distance}
\label{ssec:dL}
Galaxy samples are often defined by its observed flux or the rest-frame
luminosity inferred from the observed flux. The fluctuation in the luminosity 
distance at the observed redshift~$\zz$ is defined as
\beeq
\label{eq:dL}
\dL(\zz)\equiv\bar D_L(\zz)(1+\ddL)~,\qquad
\bar D_L(\zz)=(1+\zz)\rbar(\zz)~,
\eneq
where the fluctuation~$\ddL$ is dimensionless. Noting that the luminosity
distance is related to the angular diameter distance
\beeq
\label{eq:dA}
\dA(\zz)={\dL(\zz)\over(1+\zz)^2}~,
\eneq
we can utilize the calculations of the photon path measured by the observer
at origin to compute the angular diameter distance, and the fluctuation in
the angular diameter distance is identical to the fluctuation in the luminosity
distance. The fluctuation in the luminosity distance
has been computed in \cite{SASAK87,FUSA89,BODUGA06,JESCHI12,DIDU12}, 
and the second-order
calculations are recently presented in \cite{BEMAET12a,BEGAET13}. 
Here we briefly present the calculation, but express it in terms of
distortions in photon path we computed in Sec.~\ref{ssec:path}, which 
clearly highlights the physical effects in play.

Let's consider a unit area $dA_\up{phy}$ in the source rest frame that appears 
subtended by the observed solid angle $d\Omega=\sin\ttt d\ttt d\pp$.
This unit area is related to the angular diameter distance as
$dA_\up{phy}=\dA^2(\zz)d\Omega$, and similar to the calculation 
in Sec.~\ref{ssec:dV} it can be computed in a covariant way as
\beeq
dA_\up{phy}=\sqrt{-g}~\varepsilon_{dabc}\UU^d_s\NN^a_s
{\partial x^b_s\over\partial\ttt}{\partial x^c_s\over\partial\pp}~d\ttt~d\pp ~,
\eneq
where the velocity four vector defines the source rest-frame and the observed
photon direction defines the unit area in the source frame. The observed
photon vector in Eq.~(\ref{eq:opho}) 
\beeq
\NN^a={k^a\over k^b\UU_b}+\UU^a~
\eneq
is the observed photon direction expressed in a FRW frame and parallelly
transported along the photon path. This is not to be confused with the
observed photon direction $n^\alpha=(\ttt,\pp$) measured in the observer 
rest frame. Therefore, the angular diameter distance is
\bear
\dA^2(\zz)&=&{\sqrt{-g}~\varepsilon_{dabc}\over\sin\ttt}~\UU^d\NN^a
{\partial x^b_s\over\partial\ttt}{\partial x^c_s\over\partial\pp}
=\bar D_A^2(\zz)(1+\dg)(1+\dz)^2~\left[{\varepsilon_{dabc}\over\rbar_z^2
\sin\ttt}\left(a\UU^d_s\right)\left(a\NN^a_s\right)
{\partial x^b_s\over\partial\ttt}{\partial x^c_s\over\partial\pp}\right]\\
&=&\bar D_A^2(\zz)\left[1+2\ddL+\ddL^2\right]~.\nonumber
\enar
To simplify the calculation, we compute the square bracket by
splitting it into three components,
\beeq
\label{eq:twocomp}
{\varepsilon_{dabc}\over\rbar_z^2
\sin\ttt}\left(a\UU^d_s\right)\left(a\NN^a_s\right)
{\partial x^b_s\over\partial\ttt}{\partial x^c_s\over\partial\pp}
={\varepsilon_{0\alpha\beta\gamma}\over\rbar_z^2
\sin\ttt}\left(a\NN^\alpha_s\right)
{\partial x^\beta_s\over\partial\ttt}{\partial x^\gamma_s\over\partial\pp}+
{\varepsilon_{0\alpha\beta\gamma}\over\rbar_z^2
\sin\ttt}~\dUU^0\left(a\NN^\alpha_s\right)
{\partial x^\beta_s\over\partial\ttt}{\partial x^\gamma_s\over\partial\pp}+
{\varepsilon_{\delta abc}\over\rbar_z^2
\sin\ttt}\left(a\UU^\delta_s\right)\left(a\NN^a_s\right)
{\partial x^b_s\over\partial\ttt}{\partial x^c_s\over\partial\pp}~,
\eneq
and the third component vanishes in the mean and the linear order in
perturbations:
\beeq
{\varepsilon_{\delta abc}\over\rbar_z^2
\sin\ttt}\left(a\UU^\delta_s\right)\left(a\NN^a_s\right)
{\partial x^b_s\over\partial\ttt}{\partial x^c_s\over\partial\pp}=
-\VTT\left(\VV^\alpha-\BB^\alpha\right)e_\alpha-{1\over\rbar_z}
\left(\VTT{\partial\over\partial\ttt}+{\VPP\over\sin\ttt}{\partial\over
\partial\pp}\right)\dTau~,
\eneq
where the first term arises from the component 
$\varepsilon_{\delta0\beta\gamma}$ and the second term from 
$\varepsilon_{\delta\alpha bc}$. The second component in Eq.~(\ref{eq:twocomp})
is 
\beeq
{\varepsilon_{0\alpha\beta\gamma}\over\rbar_z^2
\sin\ttt}~\dUU^0\left(a\NN^\alpha_s\right)
{\partial x^\beta_s\over\partial\ttt}{\partial x^\gamma_s\over\partial\pp}=
\dUU^0+\dUU^0\left(-2~\kappa+2~{\drr\over\rbar_z}-\CC_{\alpha\beta}e^\alpha
e^\beta\right)~,
\eneq
and finally the first component can be computed as
\bear
&&{\varepsilon_{0\alpha\beta\gamma}\over\rbar_z^2
\sin\ttt}\left(a\NN^\alpha_s\right)
{\partial x^\beta_s\over\partial\ttt}{\partial x^\gamma_s\over\partial\pp}=
1+\VP+\eP-\dhnu+2~{\drr\over\rbar_z}+
\left(\cot\ttt+{\partial\over\partial\ttt}\right)\dtt+{\partial\over\partial
\pp}\dpp \\
&&\hspace{30pt}
-2~\kappa\left(2~{\drr\over\rbar_z}-\CC_{\alpha\beta}e^\alpha e^\beta\right)
-2~{\drr\over\rbar_z}~\CC_{\alpha\beta}e^\alpha e^\beta
+\left({\drr\over\rbar_z}\right)^2-\dtt^2-{1\over2}\left(\sin\ttt~\dpp\right)^2
+\dhnu\left(\VP+\CC_{\alpha\beta}e^\alpha e^\beta\right)\nonumber \\
&&\hspace{30pt}
-{1\over\rbar_z}\left[\left(\VTT+\eTT\right){\partial\over\partial\ttt}
+{\left(\VPP+\ePP\right)\over\sin\ttt}{\partial\over\partial\pp}\right]\drr
+\left(\VTT+\eTT\right)\dtt+\left(\VPP+\ePP\right)\sin\ttt~\dpp
\nonumber \\
&&\hspace{30pt}
-{\partial\over\partial\pp}\dtt{\partial\over\partial\ttt}\dpp
+{\partial\over\partial\pp}\dpp{\partial\over\partial\ttt}\dtt
+\cot\ttt\dtt\left({\partial\over\partial\ttt}\dtt+
{\partial\over\partial\pp}\dpp\right)
+\left(\dtt{\partial\over\partial\ttt}+\dpp{\partial\over\partial\pp}
\right){\drr\over\rbar_z}~,\nonumber
\enar
where the spatial perturbation to the photon vector at the source position
is decomposed as
\beeq
\label{eq:deadec}
\dea^\alpha\equiv\eP~n^\alpha+\eTT~\theta^\alpha+\ePP~\phi^\alpha~.
\eneq
Collecting terms altogether and using the null equation, 
the fluctuation in the luminosity distance is obtained as
\bear
\ddL^{(1)}&=&\dz-\kappa+{\drr\over\rbar_z}
+{1\over2}\left(\CC^\alpha_\alpha-\CC_{\alpha\beta}e^\alpha e^\beta\right)~, \\
2\ddL^{(2)}&=&2~\dz+\dg+\dUU^0+
\VP+e_\alpha\dea^\alpha-\dhnu+2~{\drr\over\rbar_z}+
\left(\cot\ttt+{\partial\over\partial\ttt}\right)\dtt+{\partial\over\partial
\pp}\dpp\\
&&
\dUU^0\left(2~\kappa-2~{\drr\over\rbar_z}+\CC_{\alpha\beta}e^\alpha
e^\beta\right)-\VTT\left(\VV^\alpha-\BB^\alpha\right)e_\alpha-{1\over\rbar_z}
\left(\VTT{\partial\over\partial\ttt}+{\VPP\over\sin\ttt}{\partial\over
\partial\pp}\right)\dTau~\nonumber\\
&&+\dg\left(\dz-{1\over4}~\dg\right)-\left(\AA+2~\kappa-2~{\drr\over\rbar_z}
+\CC_{\alpha\beta}e^\alpha e^\beta\right)\left[\ddL+{3\over4}
\left(\AA+2~\kappa-2~{\drr\over\rbar_z}
+\CC_{\alpha\beta}e^\alpha e^\beta\right)\right] \nonumber\\
&&
-2~\kappa\left(2~{\drr\over\rbar_z}-\CC_{\alpha\beta}e^\alpha e^\beta\right)
-2~{\drr\over\rbar_z}~\CC_{\alpha\beta}e^\alpha e^\beta
+\left({\drr\over\rbar_z}\right)^2-\dtt^2-{1\over2}\left(\sin\ttt~\dpp\right)^2
+\dhnu\left(\VP+\CC_{\alpha\beta}e^\alpha e^\beta\right)\nonumber \\
&&
-{1\over\rbar_z}\left[\left(\VTT+\eTT\right){\partial\over\partial\ttt}
+{\left(\VPP+\ePP\right)\over\sin\ttt}{\partial\over\partial\pp}\right]\drr
+\left(\VTT+\eTT\right)\dtt+\left(\VPP+\ePP\right)\sin\ttt~\dpp
\nonumber \\
&&
-{\partial\over\partial\pp}\dtt{\partial\over\partial\ttt}\dpp
+{\partial\over\partial\pp}\dpp{\partial\over\partial\ttt}\dtt
+\cot\ttt\dtt\left({\partial\over\partial\ttt}\dtt+
{\partial\over\partial\pp}\dpp\right)
+\left(\dtt{\partial\over\partial\ttt}+\dpp{\partial\over\partial\pp}
\right){\drr\over\rbar_z}+\DX^a{\ddL}_{,a}~.\nonumber
\enar
Given the distortion at the observed redshift,
the fluctuation arises due to the distortion in the solid angle,
the radial displacement, and the rest-frame of the source at the linear
order in perturbations. To the second order, in addition to the nonlinear
coupling of the linear order terms, there exist additional distortions
along the tangential directions, as was the case in the volume distortion.

\subsection{Observed galaxy number density}
\label{ssec:ng}
In observation, the galaxy number density $n_g^\up{obs}$ is obtained 
by counting the number $dN_g^\up{obs}$ of galaxies observed 
within the volume defined by the observed direction $(\ttt,\pp)$ and the
observed redshift~$\zz$: $dN_g^\up{obs}=n_g^\up{obs}dV_\up{obs}$~, where
the volume element $dV_\up{obs}$
over the small interval $(d\zz,d\ttt,d\pp)$ in observation is 
\beeq
\label{eq:obsdv}
dV_\up{obs}={\rbar_z^2\over H_z(1+\zz)^3}~\sin\ttt~d\zz ~d\ttt ~d\pp~.
\eneq
It is the volume element in a homogeneous universe based on the observed
quantities $(\zz,\ttt,\pp)$, and it is the only quantity in cosmological 
observations that can be assigned to the observed volume element in 
a physically meaningful way. This would correspond to an observer's choice
of gauge condition, {\it uniform-redshift} gauge.\footnote{The observed 
redshift is gauge-invariant, since its value (the observed
redshift) remains unchanged, whatever coordinate system is used to describe it.
However, it depends on the frame of an observer, i.e., the observed redshift
is the spectral line ratio measured by an observer at rest. When an observer
compares the real physical universe to a homogeneous universe, a choice of
gauge condition needs to be made, and it is based on the observed redshift.}

However, since the Universe is far from being homogeneous,
the constructed volume $dV_\up{obs}$ in Eq.~(\ref{eq:obsdv})
differs from the physical volume $dV_\up{phy}$ 
in Eq.~(\ref{eq:dv}) occupied by the 
observed galaxies on the sky. Using the conservation of the number of
galaxies $dN^\up{obs}_g$, the observed galaxy number density is related to
the physical number density $n_g$ of the observed source galaxies
defined in their rest frame as
\beeq
n_g^\up{obs}=n_g~(1+\dV)~.
\eneq
This relation highlights the contribution of the volume distortion~$\dV$ 
in Eq.~(\ref{eq:ddv}), and the volume effect is present in galaxy clustering,
regardless of which galaxy sample is selected \cite{YOO09}.

Furthermore, the physical number density $n_g$
of source galaxies can be separated
into the mean and the remaining fluctuation as
\beeq
\label{eq:bng}
n_g=\bar n_g(t_p)\left(1+\delta_g^\up{int}\right)~,
\eneq
where the mean is obtained by averaging the number density over a hypersurface
defined by some time coordinate~$t_p$ and the intrinsic
fluctuation around the mean vanishes when averaged:
\beeq
\bar n_g(t_p)\equiv
\langle n_g\rangle_{t_p}~,\qquad \langle\delta_g^\up{int}\rangle_{t_p}=0~.
\eneq
While the separation of the galaxy number density into the mean
and the fluctuation is completely arbitrary and gauge-dependent in 
Eq.~(\ref{eq:bng}) as it relies on a unspecified choice of time~$t_p$, 
a physically meaningful choice of time coordinate  (and hence gauge-invariant) 
can be made in relation to the biasing scheme, in which
the galaxy fluctuation~$\delta_g^\up{int}$ 
can be further related to the underlying
matter density fluctuation. To the linear order,
a proper time (and hence the notation $t_p$) 
can be chosen to provide a physical
biasing relation between the remaining fluctuation~$\delta_g^\up{int}$ and 
the underlying matter fluctuation~$\delta_m$
\cite{BASEET11,BODU11,CHLE11,JESCHI12,YOHAET12}, as
the local dynamics of galaxy formation can only be affected by the presence
of long wavelength modes through the change in the local curvature and 
the local expansion rate \cite{BASEET11}. 
Here we leave the second-order biasing 
to future work and proceed with a unspecified time coordinate (or unspecified
gauge choice) for the intrinsic galaxy fluctuation~$\delta_g^\up{int}$.

In addition to the intrinsic fluctuation of the source galaxies,
additional contribution to galaxy clustering arises from the source effect
\cite{YOO09}: The source effect describes the contributions of the
mean expressed in terms of observed quantities:
\beeq
n_g=\bar n_g(z)\left[1-e_1~\dz_{t_p}+{1\over2}e_2~\dz_{t_p}^2\right]
\left(1+\delta_g^\up{int}\right)~,
\eneq
where $\dz_{t_p}$ is the distortion in the observed redshift as in 
Eq.~(\ref{eq:dz}) but is evaluated at the time slicing specified by~$t_p$ and
two additional coefficients 
\beeq
\label{eq:eee}
e_1={d\ln\bar n_g\over d\ln(1+z)}~,\qquad
e_2=e_1+e_1^2+{de_1\over d\ln(1+z)}~,
\eneq
are called the evolution biases. Since the mean number density $\bar n_g$
here is a physical number density, even a sample with a constant comoving 
number density such as the matter density $\bar\rho_m$ would have $e_1=3$ and
$e_2=12$.

Furthermore, additional source effects will be present,
if the source galaxy sample is defined 
by other observable quantities such as the rest-frame luminosity threshold
inferred from the threshold in 
observed flux~$f_\up{obs}$. Similar to the observed volume
$dV_\up{obs}$, the luminosity distance 
$\bar D_L(\zz)=\rbar_z(1+\zz)$
in a homogeneous universe based on the observed redshift~$\zz$ is assigned 
to the source, and the inferred luminosity 
at a given observed flux~$f_\up{obs}$ is then
\beeq
\hat L=4\pi\bar D_L^2(\zz)f_\up{obs}~.
\eneq
The physical luminosity~$L$ of the source galaxies 
is related to the inferred luminosity as
\beeq
L=4\pi\dL^2(\zz)f_\up{obs}=\hat L(1+\ddL)^2~,
\eneq
where the physical luminosity distance
$\dL(\zz)\equiv\bar D_L(\zz)(1+\ddL)$.
Therefore,
the observed galaxy population defined its inferred luminosity
above a given threshold is related to the galaxy population with 
corresponding physical luminosity above the same threshold as
\beeq
n_g=\bar n_g(\hat L)\left[1-t_1~\ddL+{1\over2}~t_2~\ddL^2\right]
\left(1+\delta_g^\up{int}\right)~,
\eneq
where two additional coefficients
\beeq
\label{eq:ttt}
t_1\equiv-2~{d\ln\bar n_g\over d\ln L}~,\quad
t_2=t_1+t_1^2-2~{dt_1\over d\ln L}~,
\eneq
describe the slope and running of the luminosity function. 
When the luminosity function
$d\bar n_g\propto L^{-s}$ is well approximated by a constant slope~$s$, 
these coefficients are
\beeq
t_1=2(s-1)=5p~,\quad t_2=2(s-1)(2s-1)=5p~(5p+1)~,
\eneq
where $p=0.4(s-1)$ is the luminosity function slope in terms of magnitude
$M=\up{constant}-2.5\log_{10}(L/L_0)$.

\subsection{Observed galaxy fluctuation}
\label{ssec:delta}
Finally, by putting it altogether, the mean number density of the observed
galaxies is expressed in terms of the observed redshift~$\zz$ and 
the observed flux~$f_\up{obs}$, and the observed galaxy number density
is then decomposed of the mean and the remaining fluctuation as
\beeq
\label{eq:nobs}
n_g^\up{obs}
(\zz,\nhat)=\bar n_g(\zz)\bigg(1+\delta_g^\up{int}\bigg)\bigg(1+\dV\bigg)
\bigg(1-e_1~\dz_{t_p}+{1\over2}~e_2~\dz^2_{t_p}\bigg)
\bigg(1-t_1~\ddL+{1\over2}~t_2~\ddL^2\bigg)~.
\eneq
This equation concisely summarizes
the main result of the paper, in conjunction with
the computation of all the perturbation quantities present
in Eq.~(\ref{eq:nobs}). It is noted that
only the metric perturbations are expanded to the second order, while the
intrinsic fluctuation $\delta_g^\up{int}$ that is likely to be highly 
nonlinear is left unexpanded. Furthermore, no gauge choice is made
in the previous calculations.

In the absence of {\it ab initio} knowledge of galaxy formation, the mean 
galaxy number density $\bar n_g(z)$ cannot be computed {\it a priori} ---
it has to be determined by the survey itself. Therefore, the observed mean 
at each redshift
is obtained by averaging the observed number density~$n_g^\up{obs}(\zz,\nhat)$
over the survey area~$\Omega$:
\beeq
\widehat{\bar n_g}(\zz)\equiv{1\over\Omega}\int_\Omega d^2\nhat~n_g^\up{obs}
(\zz,\nhat)\equiv\bar n_g(z)+\delta\bar n_g(z)~,
\eneq
where the residual fluctuation $\delta\bar n_g(z)$ in the mean number density
arises if all the fluctuations in Eq.~(\ref{eq:nobs}) may not average out
over the survey area. Only in the limit of infinite volume survey, the residual
fluctuation vanishes, and we have $\widehat{\bar n_g}(\zz)=\bar n_g(\zz)$
at all redshifts. The detailed calculation of the observed mean involves
the survey specifications and the spatial distribution of fluctuations
in Eq.~(\ref{eq:nobs}) that cannot be computed with generality.
The observed galaxy fluctuation is defined in terms of the observed mean
number density as
\beeq
\label{eq:dobs0}
\delta_g^\up{obs}(\zz,\nhat)\equiv {n_g^\up{obs}
(\zz,\nhat)\over\widehat{\bar n_g}(\zz)}
-1~,
\eneq
and it is noted that the residual fluctuation $\delta\bar n_g(\zz)$ 
contributes to the observed mean number density $\widehat{\bar n_g}$
and the observed galaxy fluctuation $\delta_g^\up{obs}$.

By assuming the infinite survey volume $\widehat{\bar n_g}(\zz)=\bar n_g(\zz)$,
we derive
\bear
\label{eq:dobs1}
\delta_g^{\up{obs}(1)}&=&\delta_g^{\up{int}(1)}+\dV-e_1~\dz_{t_p}-t_1~\ddL~,\\
\label{eq:dobs2}
\delta_g^{\up{obs}(2)}&=&\delta_g^{\up{int}(2)}+\dV-e_1~\dz_{t_p}
+{1\over2}~e_2~\dz^2_{t_p}-t_1~\ddL+{1\over2}~t_2~\ddL^2-e_1~\dz_{t_p}~\dV
\nonumber \\
&&+\delta_g^\up{int}\left(\dV-e_1~\dz_{t_p}-t_1~\ddL\right)
-t_1~\ddL\left(\dV-e_1~\dz_{t_p}\right)~.
\enar

\section{Gauge Invariant Equations}
\label{sec:gie}
Having derived all the equations without choosing a gauge condition
in Sec.~\ref{sec:gcgr}, we construct the gauge-invariant equations 
for computing the observed galaxy fluctuation. Compared to the
linear-order calculations, the second-order calculations regarding gauge
transformations are more complicated, and they are further affected by
the presence of unphysical gauge modes. However, once gauge modes are
removed and gauge-invariant variables are constructed, it is straightforward
to construct second-order gauge-invariant equations, given the gauge-invariant
equations at the linear order, although  a complete
verification of second-order
gauge-invariance associated with those equations
is more involved.

In Appendix~\ref{app:gt}, we explicitly derive the second-order gauge
transformation to isolate and remove the gauge modes to the second order 
in perturbations. Using the gauge-transformation properties, 
second-order gauge-invariant variables are explicitly constructed
in Appendix~\ref{app:gi}, and their structure takes a rather simple form.
For example, the linear-order gauge-transformation 
($\tilde\tau=\tau+T$) at a given position yields the metric transformation
\beeq
\label{eq:chid}
\chi\equiv a(\beta+\gamma')~,\quad
\tilde\chi^{(1)}=\chi^{(1)}-aT^{(1)}~,\quad 
\tilde\varphi^{(1)}=\varphi^{(1)}-\HH~T^{(1)}~,
\eneq
and we can construct a linear-order gauge-invariant variable
\beeq
\label{eq:pxfirst}
\px^{(1)}\equiv \varphi^{(1)}-H\chi^{(1)}~.
\eneq
However, as we explicitly show in Eq.~(\ref{app:gtpx}),
this combination becomes gauge-dependent at the second order, and
additional compensating terms are required to cancel the second-order
corrections and guarantee its gauge-invariance. Therefore, the second-order
gauge-invariant variable can be written in a form:
\beeq
\label{eq:pxsecond}
\px=\varphi-H\chi+\px^\qqq~,
\eneq
where the last term represents quadratic terms that compensate for the
second-order gauge-transformation and its explicit expression is shown
in Eq.~(\ref{app:pxgi}). As demonstrated in Appendix~\ref{app:gt},
a choice of gauge condition $\chi=0$ to the second order in perturbations
completely removes unphysical gauge modes, and the remaining metric
perturbations correspond to gauge-invariant variables associated with
the choice of gauge condition $\chi=0$. While a variety of second-order
gauge-invariant variables can be constructed satisfying Eq.~(\ref{eq:pxfirst})
at the linear order, we constructed $\px$ in Eq.~(\ref{eq:pxsecond}),
such that $\px$ becomes $\varphi$ when the gauge condition $\chi=0$ is
adopted (hence the notation). Therefore, the quadratic terms $\px^\qqq$ in 
Eq.~(\ref{eq:pxsecond}) or in Eq.~(\ref{app:pxgi}) satisfy the relation
\beeq
\px^\qqq=0\quad\up{if}~\chi=0~,
\eneq
which greatly simplifies the way to construct second-order gauge-invariant
equations.

Before we start constructing gauge-invariant equations for those we
derived in Sec.~\ref{sec:gcgr}, we caution that not all equations can be
made gauge-invariant, but this statement should not be confused
with the fact that all equations with any proper choice of gauge
condition are gauge-invariant. The gauge-invariance of equations 
itself does not guarantee that they describe observable or physical 
quantities, but it provides a necessary condition for those equations.
We start constructing the second-order gauge-invariant equations, 
step by step, with the null equation~(\ref{eq:null}) as a worked example, 
and we then present the remaining gauge-invariant equations 
in Sec.~\ref{sec:gcgr}. 

To the linear order in perturbations, the null 
equation can be re-arranged as
\beeq
0=2\left(e_\alpha\dea^\alpha-\dnu-\AA
+\BB_\alpha~e^\alpha+\CC_{\alpha\beta}~e^\alpha e^\beta\right)^{(1)}
=2\left(e_\alpha\deag-\dnug-\ax+\px
+\pv_\alpha~e^\alpha+\TCC_{\alpha\beta}~e^\alpha e^\beta\right)^{(1)}~,
\eneq
where the definition of the gauge-invariant variables are explicitly
present in Appendix~\ref{app:gi}.
However, the above expression is not gauge-invariant to the second order
in perturbations, because of the quadratic terms in the gauge-invariant
variables and the remaining quadratic terms in the null equations.
To the second order, we re-arrange the null equation as
\bear
\label{eq:intgi}
0&=&2(e_\alpha\deag-\dnug-\ax+\px+\pv_{\chi\alpha} ~e^\alpha
+\TCC_{\chi\alpha\beta}~
e^\alpha e^\beta) -2(e_\alpha ~\deag-\dnug-\ax+\px+\pv_{\chi\alpha} ~e^\alpha
+\TCC_{\chi\alpha\beta}e^\alpha e^\beta)^\qqq \nonumber \\
&&
+\dea^\alpha~\dea_\alpha-\dnu^2-2~\dnu~\left(2\AA-\BB_\alpha e^\alpha\right)
+2\left(\BB_\alpha+2~\CC_{\alpha\beta}~e^\beta\right)\dea^\alpha~,
\enar
and it is noted that a choice of scalar gauge condition is needed to
construct second-order gauge-invariant vectors and tensors. Since we eliminate
the unphysical gauge modes to the second-order in perturbations by choosing
the spatial C-gauge condition in Appendix~\ref{app:gt}, the metric tensor then 
corresponds to
\beeq
\AA = \alpha, \quad 
\BB_\alpha = {1 \over a} \chi_{,\alpha}+\pv_\alpha, \quad
\CC_{\alpha\beta} = \varphi~\gbar_{\alpha\beta}+\TCC_{\alpha\beta}~.
\eneq
Therefore, the remaining quadratic terms in Eq.~(\ref{eq:intgi}) can be
readily re-arranged, and the gauge-invariant equation for the null condition
becomes
\bear
0&=&\CK^a\CK_a=\left(e^\alpha e_\alpha-1\right)
+2\left(e_\alpha\deag-\dnug-\ax+\px
+\pv_{\chi\alpha}~e^\alpha+\TCC_{\chi\alpha\beta}~e^\alpha e^\beta\right)
\nonumber \\
&&+\deag\dea_{\chi\alpha}-\dnug^2
-2\dnug\left(2\ax-\pv_{\chi\alpha} e^\alpha\right)
+2\left(2\px e_\alpha+\pv_{\chi\alpha}
+2~\TCC_{\chi\alpha\beta}~e^\beta\right)\deag~.
\enar

Construction of gauge-invariant expressions for geodesic equations can be
made in a similar way by noting that the affine parameter integration
deviates from the straight line at the second order.
The temporal component of the geodesic equation is
\bear
-{d\over d\cc}\dnug&=&\ax'+\px'-2{\ax}_{,\alpha}e^\alpha
+\left(\pv_{\chi\alpha|\beta}+\TCC_{\chi\alpha\beta}'\right) e^\alpha e^\beta
+2\dnug\left(\ax' - {\ax}_{,\alpha}e^\alpha\right)
-2\ax\ax' -{\ax}_{,\alpha}\pv^\alpha+\pv^\alpha\pv_\alpha'\nonumber \\
&&
-2\ax\px'-2{\px}_{,\alpha}e^\alpha\pv_\alpha e^\alpha+\px^{,\alpha}\pv_{\alpha}
- \left[\ax\left( 2\pv_{\alpha|\beta} 
+2\TCC_{\alpha\beta}'\right)+\pv_\gamma \left( 2 \TCC^\gamma_{\alpha|\beta}
-\TCC_{\alpha\beta}^{\;\;\;\;|\gamma}\right)\right]e^\alpha e^\beta
 \nonumber\\
&&
+2\left(2\ax{\ax}_{,\alpha}+\px'\pv_\alpha+\pv_\beta\TCC_\alpha^{\beta\prime}
+\pv^\beta \pv_{[\alpha|\beta]} \right) e^\alpha
-2\deag\left[{\ax}_{,\alpha}-\px' e_\alpha
-e^\beta\left(\pv_{(\alpha|\beta)} +\TCC_{\alpha\beta}'\right)\right]
~,
\enar
and the spatial component is
\bear
{d\over d\cc}~\deag&=&\ax^{\;\;,\alpha}-\px^{\;\;,\alpha}
-2\left(\px'-{\px}_{,\beta}e^\beta\right)e^\alpha
-\pv_\chi^{\alpha\prime}
- \left( \pv_{\chi\beta}^{\;\;\;|\alpha} - \pv^\alpha_{\chi|\beta}
+ 2 \TCC^{\alpha\prime}_\beta \right) e^\beta+ 
\left( 2 \TCC^\alpha_{\beta|\gamma} - \TCC_{\beta\gamma}^{\;\;\;\;|\alpha}
\right) e^\beta e^\gamma\nonumber \\
&&
+2\dnug\left(\ax^{\;\;,\alpha} - \pv^{\alpha\prime} \right)
-\left( \dea_\chi^\beta + \dnug e^\beta \right)\left(2\px'\delta^\alpha_\beta+
\pv_\beta^{\;\;|\alpha} 
- \pv^\alpha_{\;\;|\beta}+ 2 \TCC^{\alpha\prime}_\beta \right)
+ 2 \left( 2 \TCC^\alpha_{\beta|\gamma}-\TCC_{\beta\gamma}^{\;\;\;\;|\alpha} 
\right)e^\beta \dea_\chi^\gamma \nonumber \\
&&+4~{\px}_{,\beta}\dea^\beta_\chi~e^\alpha-2~e_\beta\dea^\beta_\chi
\px^{\;\;,\alpha}+ \ax' \pv^\alpha-2~\px\ax^{\;\;,\alpha}
+2~\px\pv^{\alpha\prime}- 2 {\ax}_{,\beta} \TCC^{\alpha\beta}
+ 2 \TCC^\alpha_\beta \pv^{\beta\prime}\nonumber \\
&&
- 2 \pv^\alpha {\ax}_{,\beta} e^\beta+
4\left(\px\px'\delta^\alpha_\beta+\px\TCC^{\prime\alpha}_\beta
+\pv_{[\beta|\gamma]}\TCC^{\alpha\gamma}+\px'\TCC^\alpha_\beta
+\TCC^{\alpha\gamma}\TCC'_{\beta\gamma}\right)e^\beta
+\px'\pv^\alpha+\pv^\alpha \left( \pv_{\beta|\gamma} + \TCC'_{\beta\gamma}
\right) e^\beta e^\gamma  
 \nonumber\\
&&
- 2~\left(\px\delta^\alpha_\delta+ \TCC^\alpha_\delta\right)
\left( 2{\px}_{,\gamma}\delta^\delta_\beta+2\TCC^\delta_{\beta|\gamma}
- \px^{,\delta}\gbar_{\beta\gamma}-
\TCC_{\beta\gamma}^{\;\;\;\; |\delta} \right)  e^\beta e^\gamma  ~.
\enar
Integrating the geodesic equation over the affine parameter, we derive
temporal and spatial deviations
\bear
\dnug\bigg|_o^\zz&=&-2\left({\ax}_z-{\ax}_o\right)
-\int_0^{\rbar_z}d\rbar~\bigg\{\ax'-\px'
-\left(\pv_{\chi\alpha|\beta}+\TCC_{\chi\alpha\beta}'\right) e^\alpha e^\beta
+2\dnug {\ax}_{,\alpha}e^\alpha+2\ax\ax' +{\ax}_{,\alpha}\pv^\alpha
\nonumber \\
&&
-\pv^\alpha\pv_\alpha'
+2\ax\px'+2{\px}_{,\alpha}e^\alpha\pv_\alpha e^\alpha-\px^{,\alpha}\pv_{\alpha}
+ \left[\ax\left( 2\pv_{\alpha|\beta} 
+2\TCC_{\alpha\beta}'\right)+\pv_\gamma \left( 2 \TCC^\gamma_{\alpha|\beta}
-\TCC_{\alpha\beta}^{\;\;\;\;|\gamma}\right)\right]e^\alpha e^\beta
 \nonumber\\
&&
-2\left(2\ax{\ax}_{,\alpha}+\px'\pv_\alpha+\pv_\beta\TCC_\alpha^{\beta\prime}
+\pv^\beta \pv_{[\alpha|\beta]} \right) e^\alpha
-2\deag\left[\px' e_\alpha
+e^\beta\left(\pv_{(\alpha|\beta)} +\TCC_{\alpha\beta}'\right)\right]
\nonumber \\
&&
+\DX_\chi^c\left[\ax'-\px'
-\left(\pv_{\chi\alpha|\beta}+\TCC_{\chi\alpha\beta}'\right) e^\alpha e^\beta
\right]_{,c}\bigg\}~, \\
\deag\bigg|_o^\zz&=&
-\bigg[2\px e^\alpha+\pv_\chi^\alpha+2\TCC^\alpha_{\chi\beta} 
e^\beta\bigg]^\zz_o-\int_0^{\rbar_z} d\rbar\bigg\{
\left(\ax-\px- \pv_{\chi\beta}  e^\beta
- \TCC_{\beta\gamma} e^\beta e^\gamma\right)^{|\alpha}
+\dnug\left(2\ax^{\;\;,\alpha} - \pv^{\alpha\prime} \right)
\nonumber \\
&&
-\left( \dea_\chi^\beta + \dnug e^\beta \right)\pv_\beta^{\;\;|\alpha} 
+\dnug~\pv^\alpha_{\;\;|\beta}e^\beta-2\px'\dea^\alpha_\chi
-2\TCC^{\alpha\prime}_\beta\dea^\beta_\chi
+ 2 \left(  \TCC^\alpha_{\beta|\gamma}-\TCC_{\beta\gamma}^{\;\;\;\;|\alpha} 
\right)e^\beta \dea_\chi^\gamma \nonumber \\
&&+2~{\px}_{,\beta}\dea^\beta_\chi~e^\alpha-2~e_\beta\dea^\beta_\chi
\px^{\;\;,\alpha}+ \ax' \pv^\alpha-2~\px\ax^{\;\;,\alpha}
+2~\px\pv^{\alpha\prime}- 2 {\ax}_{,\beta} \TCC^{\alpha\beta}
+ 2 \TCC^\alpha_\beta \pv^{\beta\prime}\nonumber \\
&&
- 2 \pv^\alpha {\ax}_{,\beta} e^\beta+
4\left(\px\px'\delta^\alpha_\beta+\px\TCC^{\prime\alpha}_\beta
+\pv_{[\beta|\gamma]}\TCC^{\alpha\gamma}+\px'\TCC^\alpha_\beta
+\TCC^{\alpha\gamma}\TCC'_{\beta\gamma}\right)e^\beta
+\px'\pv^\alpha \nonumber\\
&&
+\pv^\alpha \left( \pv_{\beta|\gamma} + \TCC'_{\beta\gamma}
\right) e^\beta e^\gamma  
- 2~\left(\px\delta^\alpha_\delta+ \TCC^\alpha_\delta\right)
\left( 2{\px}_{,\gamma}\delta^\delta_\beta+2\TCC^\delta_{\beta|\gamma}
- \px^{,\delta}\gbar_{\beta\gamma}-
\TCC_{\beta\gamma}^{\;\;\;\; |\delta} \right)  e^\beta e^\gamma  
\nonumber \\
&&+\DX^d_\chi\left(\ax^{\;\;,\alpha}-\px^{\;\;,\alpha} 
-\pv_{\chi\beta}^{\;\;|\alpha} e^\beta
- \TCC_{\beta\gamma}^{\;\;\;\;|\alpha}e^\beta e^\gamma\right)_{,d}\bigg\}~.
\enar
Using the above expressions, the distortion in the observed redshift can
be written in a gauge-invariant form as
\bear
\dzg^{(1)}&=&\DA_o+\bigg[\dnug+ \ax +\left( \VV_\alpha - \pv_\alpha\right)
e^\alpha\bigg]_o^\zz=\DA_o+\bigg[\left( \VV_\alpha - \pv_\alpha\right)
e^\alpha-\ax\bigg]_o^\zz-\int_0^{\rbar_z}d\rbar~\bigg[(\ax-\px)'
-\left(\pv_{\alpha|\beta}+\TCC_{\alpha\beta}'\right) e^\alpha e^\beta
\bigg]~,\nonumber\\
\dz^{(2)}_\chi&=&\DA_o+\Bigg[\dnug+ \ax +\left( \VV_\alpha - \pv_\alpha\right)
e^\alpha+\dnug\ax + \deag \left(\VV_\alpha - \pv_\alpha \right)
+ \left(2\px\VV_\alpha+ 
\ax \pv_\alpha + 2 \TCC_{\alpha\beta} \VV^\beta \right) e^\alpha
+ {1 \over 2} \VV^\alpha\VV_\alpha - {1 \over 2} \ax^2 \Bigg]^z_o \nonumber\\
&&
+\bigg[\DA-\dnug- \ax -\left( \VV_\alpha - \pv_\alpha\right)e^\alpha\bigg]_o
\Bigg[\dnug+ \ax +\left( \VV_\alpha - \pv_\alpha\right)e^\alpha\Bigg]^z_o 
+\DX_z^b\bigg[\dnug+ \ax +\left( \VV_\alpha - \pv_\alpha\right)e^\alpha
\bigg]_{,b}~.
\enar
Finally, the spatial deviations of the observed source position can be
expressed as
\bear
\label{eq:drg}
\drr_\chi^{(1)}&=&\dT_o-{\dzg\over\HH_z}+\int_0^{\rbar_z}d\rbar~\left(
\ax-\px-\pv_\alpha e^\alpha-\TCC_{\alpha\beta}e^\alpha e^\beta \right)~,\\
\rbar_z\dtt_\chi^{(1)}
&=&\rbar_ze_{\theta\alpha}\left(\deag+\pv^\alpha+2~\TCC^\alpha_\beta
e^\beta\right)_o-\int_0^{\rbar_z}d\rbar~\left[e_{\theta\alpha}
\left(\pv^\alpha+2~\TCC^\alpha_\beta e^\beta\right)+\left({\rbar_z-\rbar\over
\rbar}\right){\partial\over\partial\ttt}
\left(\ax-\px-\pv_\alpha e^\alpha-\TCC_{\alpha\beta}e^\alpha e^\beta \right)
\right]~,\nonumber
\enar
where the azimuthal deviation $\dpp_\chi^{(1)}$ can be readily inferred
from $\dtt_\chi^{(1)}$. To the second order in perturbations, they are further
related as
\bear
(n_\alpha x^\alpha_s)^{(2)}_\chi
&=&\dT_o^{(2)}+\left(\dT_z-{\dz\over\HH_z}\right)_\chi
\left(\ax-\px-\pv_\alpha e^\alpha-
\TCC_{\alpha\beta}e^\alpha e^\beta\right)_\zz
+{1\over2\HH_z^3}\left(\HH_z^2+\HH_z'\right)\dzg^2-{\dzg^{(2)}\over\HH_z}
 \\
&&+\int_0^{\rbar_z}d\rbar~\bigg\{
\left(\ax-\px-\pv_{\chi\alpha} e^\alpha-\TCC_{\chi\alpha\beta}~
e^\alpha e^\beta\right)^{(2)}
+\dnug\bigg[{1\over2}\dnug+\left(2\ax-\pv_\alpha e^\alpha\right)\bigg]
\nonumber \\
&&\hspace{50pt}
-\bigg[{1\over2}\dea_{\chi\alpha}+2\px e_\alpha
+\left(\pv_\alpha+2~\TCC_{\alpha\beta}~e^\beta\right)\bigg]\deag
+\DX^c_\chi\left(\ax-\px-\pv_\alpha e^\alpha
-\TCC_{\alpha\beta}~e^\alpha e^\beta\right)_{,c}\bigg\}
~,\nonumber \\
\label{eq:dtg}
(\ttt_\alpha x^\alpha_s)^{(2)}_\chi
&=&\rbar_z e_{\ttt\alpha}\left(\deag+\pv_\chi^\alpha
+2~\TCC^\alpha_{\chi\beta} e^\beta\right)_o^{(2)}
-\Dcc_s e_{\ttt\alpha}\dea^\alpha_z \\
&&
-\int_0^{\rbar_z}d\rbar~\bigg\{e_{\ttt\alpha}
\left(1+\DX_\chi^b\partial_b\right)
\left(\pv^\alpha+2~\TCC^\alpha_\beta e^\beta\right) 
+\left({\rbar_z-\rbar\over\rbar}\right)
{\partial\over\partial\ttt}\bigg[(1+\DX^b_\chi\partial_b)
\left(\ax-\px-\pv_\alpha e^\alpha-\TCC_{\alpha\beta}e^\alpha e^\beta\right) 
\bigg]\nonumber \\
&&
+(\rbar_z-\rbar)e_{\ttt\alpha}\bigg[
\dnug\left(2\ax^{\;\;,\alpha} - \pv^{\alpha\prime} \right)
-\pv_\beta^{\;\;|\alpha}\left(\dea_\chi^\beta + \dnug ~e^\beta\right)
-2\px'\deag-2\px^{\;\;,\alpha}e_\beta\dea^\beta_\chi
-2\ax^{\;\;,\alpha}\px
\nonumber \\
&&
+ \dnug\pv^\alpha_{\;\;|\beta}e^\beta- 2 \TCC^{\alpha\prime}_\beta\dea^\beta
+ 2\left( \TCC^\alpha_{\beta|\gamma}-\TCC_{\beta\gamma}^{\;\;\;\;|\alpha} 
\right)e^\beta \dea^\gamma + \ax' \pv^\alpha-2{\ax}_{,\beta} \TCC^{\alpha\beta}
+ 2 \TCC^\alpha_\beta \pv^{\beta\prime}- 2 \pv^\alpha {\ax}_{,\beta} e^\beta
\nonumber\\
&&
+ 4\left(\px\gbar^{\alpha\gamma}+ \TCC^{\alpha\gamma}\right) 
\left( \px'\gbar_{\beta\gamma}+\pv_{[\beta|\gamma]}
+ \TCC_{\beta\gamma}' \right) e^\beta
- \Big[2\left(\px\delta^\alpha_\delta+ \TCC^\alpha_\delta\right) 
\left( 2{\px}_{,\gamma}\delta^\delta_\beta+2\TCC^\delta_{\beta|\gamma}
-\px^{\;\;,\delta}\gbar_{\beta\gamma}- 
\TCC_{\beta\gamma}^{\;\;\;\; |\delta} \right) \nonumber \\
&&
- \pv^\alpha \left(\px'\gbar_{\beta\gamma}+\pv_{\beta|\gamma} + 
\TCC'_{\beta\gamma}\right)\Big] e^\beta e^\gamma \bigg]\bigg\}~.\nonumber 
\enar
Subsequent calculations in Sec.~\ref{ssec:kappa}$-$\ref{ssec:delta} can 
be further expressed in terms of $\drr_\chi$, $\dtt_\chi$, $\dpp_\chi$,
$\dzg$ and other gauge-invariant variables in Appendix~\ref{app:gif}.

\section{Discussion}
\label{sec:discussion}
We have extended the calculation of the general relativistic description
of galaxy clustering \cite{YOFIZA09} to the second order in metric 
perturbations without assuming any gauge conditions or adopting any 
restrictions on vector and tensor perturbations. On large scales,
metric perturbations along the photon path 
affect the photon propagation, and these subtle relativistic effects need 
to be properly taken into account in considering the relation of
the observable quantities such as the observed redshift and the angular
position of source galaxies to the physical quantities of the source galaxies.
In the past few years, linear-order relativistic effect in galaxy clustering 
has been computed \cite{YOFIZA09,YOO10,BODU11,CHLE11,BASEET11,
JESCHI12,BRCRET12,YODE13}, and it was shown \cite{YOHAET12}
that these subtle relativistic effects can be used to test general relativity
and probe the early Universe in current and future galaxy surveys.
Drawing on these previous works, we have computed the second-order relativistic
effect in galaxy clustering, an essential tool for going beyond the 
power spectrum in the era of precision measurements of galaxy clustering.

Compared to the linear-order calculations, second-order calculations are 
more involved as the interchangeability between configuration and Fourier
spaces is lost and the nonlinear coupling of the linear-order terms result
in numerous additional terms
(see, e.g., \cite{ACBAET03,NOHW04,MAWA09,BAKOET04} for reviews).
Furthermore, scalar, vector, and tensor modes of perturbation variables 
become tangled as the nonlinear coupling of the linear-order terms
source each component and affect their spatial transformation properties.
To the second-order in
perturbations, we have computed the transformation
of the metric perturbations and removed the unphysical spatial gauge-modes.
This procedure is necessary for the explicit construction of second-order 
gauge-invariant variables. As is often the case in many second-order
calculations, one may assume no vector 
or tensor at the linear order and focus on scalar modes, because in this
case no vector or tensor contributes to the scalar modes even at the second
order, simplifying the situation.
However, generation of vector and tensor is inevitable at the second
order, and the observable quantities receive contributions from
perturbations of all types, regardless of our calculational convenience. 
Hence we have constructed the second-order
gauge-invariant variables with full generality on vector and tensor 
perturbations.

It is well-known \cite{SAWO67} that the observed redshift $z_\up{obs}$ 
is different from the redshift parameter~$z_h$ 
in a homogeneous universe, because perturbations along the photon path
such as the peculiar velocity and the gravitational potential contribute
to the fluctuation $\dz$ in the observed redshift:
$1+z_\up{obs}=(1+z_h)(1+\dz)$ in Eq.~(\ref{eq:dz}), where $1+z_h=1/a$.
For exactly the same reason, the observed position $n^\alpha$
of the source galaxy on the sky is different from the position $e^\alpha$
in a homogeneous universe: $n^\alpha=e^\alpha+\dnn^\alpha$
in Eq.~(\ref{eq:fulln}), where two unit directional vectors can be obtained
from the photon wavevector $k^a$ in each case. In Appendix~\ref{app:photon},
we have derived the relation between two unit directional vectors by
explicitly constructing the physical photon wavevector in terms of the 
local observable quantities. This relation clarifies how
the additional degree of freedom supplied by the conformal transformation
in Eq.~(\ref{eq:conf}) can be {\it properly} chosen to eliminate the distortion
$\dnn^\alpha$ in the observed position of the source galaxy.

With these issues resolved, it becomes rather straightforward, albeit lengthy,
to extend the linear-order relativistic calculations to the second order.
Compared to the inferred source position 
$\hat x^a_s=(\bar\tau_z~,~\rbar_z \nhat)
=(\bar\tau_z~,~\rbar_z\sin\ttt\cos\pp~,~\rbar_z\sin\ttt\sin\pp~,
~\rbar_z\cos\ttt)$ in 
Eq.~(\ref{eq:inferred}) based on the observed redshift~$\zz$ and 
angle $(\ttt,\pp)$, the physical source position can be parametrized 
in terms of the displacements $(\dTau,~\drr,~\dtt,~\dpp)$ 
in Eq.~(\ref{eq:displ}) to all orders in perturbations, and the volume effect
in Eq.~(\ref{eq:ddv}) can be readily computed to the desired orders
in perturbations,  although
this separation of spatial and time components is gauge-dependent and
these displacements need to be further related in terms of metric 
perturbations. Compared to the linear-order volume effect in 
Eq.~(\ref{eq:ddv1}), the notable difference in the second-order volume effect
in Eq.~(\ref{eq:ddv2}) is the contribution of the tangential velocity
$(\VTT,\VPP)$, and the displacement in the time coordinate $\dTau$ of
the source galaxies, as in the transverse Doppler effect.

Finally, 
by using the second-order gauge-invariant variables, we have constructed the
second-order gauge-invariant equations for the displacements. This step
is necessary for numerically computing the displacements and the observed
galaxy number density. To the second order, as quadratic terms are present
in both the dynamical equations and the gauge-invariant variables,
a proper choice of gauge-invariant variables is essential for simplifying the
second-order gauge-invariant equations. An explicit construction 
of the second-order gauge-invariant equations was given
in Sec.~\ref{sec:gie}.

The second-order relativistic description of galaxy clustering in this work
provides the most accurate and complete description of galaxy clustering
on large scales. While it is a step forward in the era of precision cosmology,
proper applications of our second-order formalism to observations will
require several steps beyond the scope of current investigation.
First and foremost is galaxy bias to the second order in perturbations.
Irrespective of nonlinear biasing schemes, 
computation of the second-order matter density fluctuation is necessary 
even for the simplest linear biasing. A physical choice of time slicing
for the matter density should be carefully examined for galaxy bias such as in 
\cite{BASEET11}. Second, we need to compute the three-point correlation 
function and the bispectrum and to forecast the detectability of these 
statistics in future galaxy surveys. As they involve large-scale modes, 
more sophisticated methods that go beyond the distant-observer 
approximation will be needed as in \cite{DAKAJE12,DAJEKA13}.
Measurements of the three-point correlation function or the bispectrum
in future surveys would not only complement
the existing constraints from the two-point statistics, but also provide
new ways to test the general relativity and probe the early universe
through the subtle relativistic effect in galaxy clustering.

\acknowledgments
We acknowledge useful discussions with Jinn-Ouk Gong and Jai-chan Hwang.
J.~Y. is supported by the Swiss National Science Foundation 
and the Tomalla foundation grants. M.~Z. is supported
in part by NSF Grants No. PHY-0855425, and No. AST-0907969, PHY-1213563 
and by the David and Lucile Packard Foundation.
During the review process of this paper, we acknowledge communication with
Daniele Bertacca on the recent submission \cite{BEMACL14} on the same 
subject.

\vfill

\bibliography{bispec.bbl}

\appendix

\input{appendix.tex}

\input{table.tex}

\end{document}

%% file: appendix.tex
\section{Second-Order Gauge-Invariant Formalism}
\label{app:gif}
Here we present our notation for second-order gauge-invariant formalism
and discuss the gauge transformation properties in comparison to its 
linear-order counterparts. We then explicitly construct second-order
gauge-invariant variables.

\subsection{Spacetime metric perturbations}
\label{app:con}
We describe the background for a spatially homogeneous and isotropic universe
with a metric
\beeq
\label{eq:bgmetric}
ds^2=g_{ab}dx^adx^b=-a^2(\tau)~d\tau^2+a^2(\tau)~\gbar_{\alpha\beta}~dx^\alpha
dx^\beta~,
\eneq
where $a(\tau)$ is the scale factor and $\gbar_{\alpha\beta}$ is the metric 
tensor for a three-space with a constant spatial curvature 
$K=-H_0^2~(1-\Omega_\up{tot})$. To describe the real (inhomogeneous) universe,
we parametrize the perturbations to the homogeneous background metric as
\beeq
\label{eq:abc}
\delta g_{00}=-2~a^2\AA~, \quad
\delta g_{0\alpha}=-a^2\BB_\alpha ~, \quad
\delta g_{\alpha\beta}=2~a^2\CC_{\alpha\beta}~.
\eneq
These perturbation variables are defined in a non-perturbative way, such
that they contain higher-order terms. To the second-order in perturbations, 
we may explicitly split the variables based on the perturbation order 
represented by the upper indices as
\beeq
\AA=\AA^{(1)}+\AA^{(2)}~,\quad \BB_\alpha=\BB^{(1)}_\alpha+\BB^{(2)}_\alpha~,
\quad \CC_{\alpha\beta}=\CC^{(1)}_{\alpha\beta}+\CC^{(2)}_{\alpha\beta}~.
\label{eq:mpert}
\eneq
Unless otherwise explicitly indicated, other perturbation variables should
also be considered as those
variables with higher-order terms. 

It is customary in cosmological perturbation theory to decompose perturbation
variables into scalar, vector, and tensor solutions of the generalized
Helmholz equation \cite{BARDE80}, according to their spatial-coordinate 
transformation properties. Therefore, the metric perturbations in
Eq.~(\ref{eq:mpert}) are further decomposed as
\beeq
\label{eq:decmet}
\AA=\alpha~,\quad \BB_\alpha=\beta_{,\alpha}+\VBB_\alpha~,\quad
\CC_{\alpha\beta}=\varphi~\gbar_{\alpha\beta}+\gamma_{,\alpha|\beta}+
\VCC_{(\alpha|\beta)}+\TCC_{\alpha\beta}~,
\eneq
where the vertical bar represents the covariant derivative with respect to
the homogeneous spatial metric $\gbar_{\alpha\beta}$ and the round bracket
is the symmetrization symbol. Separation of scalar, vector, and tensor
can be readily made, based on the number of their spatial indices in the
decomposed fields. The decomposed scalar perturbations can be obtained as
\bear
\label{eq:decomp1}
\alpha&=&\AA~,\quad \beta=\Delta^{-1} \nabla^\alpha\BB_\alpha~,  \quad
\gamma={1\over2}\left(\Delta+{1\over2}\Rbar\right)^{-1}
\left(3\Delta^{-1}\nabla^\alpha\nabla^\beta\CC_{\alpha\beta}
-\CC_{\alpha}^{\alpha} \right)~,   \\
\varphi&=&{1\over3}\CC_{\alpha}^{\alpha}-{1\over6}\Delta\left(\Delta
+{1\over2}\Rbar\right)^{-1}\left(3\Delta^{-1}\nabla^\alpha\nabla^\beta 
\CC_{\alpha\beta}-\CC_{\alpha}^{\alpha} \right)~, \nonumber
\enar
where $\nabla_\alpha$ is the covariant derivative based on 
$\gbar_{\alpha\beta}$ (i.e., vertical bar)
 and $\Delta=\nabla^\alpha\nabla_\alpha$ is the Laplacian
operator. The presence of the Ricci scalar ($\Rbar=6K$)
for the three-space indicates
that covariant derivatives are non-commutative.
The decomposed
vector and tensor components are computed in a similar manner as
\bear
\label{eq:decomp2}
\VBB_\alpha&=&\BB_\alpha-\nabla_\alpha\Delta^{-1}\nabla^\beta\BB_\beta~,
\quad
\VCC_\alpha=2\left(\Delta+{1\over3}\Rbar\right)^{-1}\left[\nabla^\beta
\CC_{\alpha\beta}-\nabla_\alpha\Delta^{-1}\nabla^\beta\nabla^\gamma
\CC_{\beta\gamma}\right]~, \\
\TCC_{\alpha\beta}&=&\CC_{\alpha\beta}-{1\over3}\CC^\gamma_\gamma
\gbar_{\alpha\beta}-{1\over2}\left(\nabla_\alpha\nabla_\beta-{1\over3}
\gbar_{\alpha\beta}\Delta\right)\left(\Delta+{1\over2}\Rbar\right)^{-1}
\left[3\Delta^{-1}\nabla^\gamma\nabla^\delta\CC_{\gamma\delta}-
\CC^\gamma_\gamma\right]~,\nonumber\\
&&-2\nabla_{(\alpha}\left(\Delta+{1\over3}\Rbar\right)^{-1}\left[
\nabla^\gamma\CC_{\beta)\gamma}-\nabla_{\beta)}\Delta^{-1}\nabla^\gamma
\nabla^\delta\CC_{\gamma\delta}\right]~,\nonumber
\enar
and they satisfy the transverse condition
$\VBB^\alpha_{\;\;\;|\alpha}=\VCC^\alpha_{\;\;\;|\alpha}
=\TCC^\beta_{\alpha|\beta}=0$ and the traceless condition 
$\TCC^\alpha_\alpha=0$. 
Note that
these variables $(\alpha,\beta,\gamma,\varphi,\VBB_\alpha,\VCC_\alpha,
\TCC_{\alpha\beta})$ are again non-linear perturbation variables, but
Eqs.~(\ref{eq:decomp1}) and~(\ref{eq:decomp2}) show that the S-V-T
decomposition is always possible in a non-perturbative way.

\subsection{Second-order gauge transformation}
\label{app:gt}
A  coordinate transformation in general relativity accompanies a transformation
of the metric tensor $g_{ab}$ and affects its correspondence of a coordinate
position to the homogeneous background universe, called a gauge transformation.
Thus, it is necessary to separate the physical degree-of-freedom from
fictitious gauge freedoms due to coordinate transformation.
Here we consider the most general coordinate 
transformation to the second order,
\beeq
\tilde x^a=x^a+\xi^a~,
\label{eq:gt}
\eneq
and decompose the infinitesimal transformation $\xi^a$
into scalar parts $T,L$ and a vector part $L^\alpha$ 
based on $\gbar_{\alpha\beta}$ as
\beeq
\label{eq:gtdec}
\xi^a=(T,~\LL^\alpha)=(T,~L^{,\alpha}+L^\alpha)~.
\eneq
While the gauge-transformation of general tensors can be
derived in terms of the Lie derivatives \cite{ABBRMU97}, we simply use the
tensor transformation properties induced by the coordinate transformation
\beeq
\tilde g_{ab}(\tilde x^e)={\partial x^c\over\partial\tilde x^a}
{\partial x^d\over\partial\tilde x^b}~g_{cd}(x^e)~,
\label{eq:gtmetric}
\eneq
where they are evaluated at the same spacetime position, represented by
two different values of coordinate components. 
Evaluating $\tilde g_{ab}$ in Eq.~(\ref{eq:gtmetric}) at $x^e$ and
relating to $g_{ab}(x^e)$, we derive
the transformation of the metric perturbations in Eq.~(\ref{eq:abc}) as
\bear
\label{eq:gtaa} 
\tilde\AA&=& \AA-\left(T^{\prime} +\HH T \right)- \AA^\prime T
- 2 \AA \left(T'+\HH T\right)+{3\over2}T^{\prime}T^{\prime}
+TT^{\prime\prime} + 3\HH T^{\prime}T+{1 \over 2} 
\left( 2\HH^2+\HH'\right) T^2  \\
&&
-\AA_{,\alpha}\LL^\alpha-\BB_\alpha\LL^{\alpha\prime}
+T_{,\alpha}\LL^{\alpha\prime}
+\LL^\alpha \left(T^{\prime}_{,\alpha}+\HH T_{,\alpha} \right)
- {1\over 2}\LL^{\alpha\prime} \LL^\prime_{\alpha}~,   \nonumber
\\
\label{eq:gtbb} 
\tilde\BB_\alpha&=&\BB_\alpha -T_{,\alpha}
- 2\AA T_{,\alpha}- \left(\BB^\prime_\alpha + 2 \HH\BB_\alpha \right) T
- \BB_\alpha T^{\prime}+2 T^{\prime} T_{,\alpha}
+T \left(T^{\prime}_{,\alpha}+ 2\HH T_{,\alpha} \right)   \\
&&+ \LL^\prime_\alpha 
-\BB_{\alpha,\beta}\LL^\beta-\BB_\beta 
\LL^\beta_{\;\;,\alpha}+ 2\CC_{\alpha\beta} \LL^{\beta\prime}  
-\LL^\prime_\alpha T^{\prime}
+ T_{,\beta}\LL^\beta_{\;\;,\alpha}+\LL^\gamma T_{,\alpha\gamma}
-T \left( \LL^{\prime\prime}_\alpha+2\HH\LL^\prime_\alpha \right)   
\nonumber \\ 
&&-\LL^{\beta}_{\;\;,\alpha} \LL^\prime_\beta-\gbar_{\alpha\beta} 
\LL^\beta_{\;\;,\gamma}\LL^{\gamma\prime}-\LL^\gamma 
\left(\gbar_{\alpha\beta,\gamma}\LL^{\beta\prime}+\gbar_{\alpha\beta} 
\LL^{\beta\prime}_{\;\;,\gamma} \right)~,   \nonumber
\\
\label{eq:gtcc}
\widetilde{\CC}_{\alpha\beta} &=& \CC_{\alpha\beta}-\HH T\gbar_{\alpha\beta}
+\BB_{(\alpha} T_{,\beta)}-\left(\CC^\prime_{\alpha\beta}+ 
2 \HH\CC_{\alpha\beta} \right)T-{1\over 2}T_{,\alpha}T_{,\beta}
+ \HH\gbar_{\alpha\beta}T^{\prime}T+ {1 \over 2} \left( 
2\HH^2+\HH'\right) \gbar_{\alpha\beta} T^2   \\
&&- {1\over 2}\gbar_{\alpha\beta,\gamma} \LL^\gamma -\gbar_{\gamma(\alpha} 
\LL^\gamma_{\;\;,\beta)}    -\CC_{\alpha\beta,\gamma} \LL^\gamma
- 2\CC_{\gamma(\alpha} \LL^\gamma_{\;\;,\beta)} + \LL^\prime_{(\alpha} 
T_{,\beta)}+\HH\gbar_{\alpha\beta}\LL^\gamma T_{,\gamma}   \nonumber \\
&&+ \left( {1\over 2}\LL^{\gamma\prime}
+\HH\LL^\gamma \right)\gbar_{\alpha\beta,\gamma}T+ 2\HH\gbar_{\gamma(\alpha}
\LL^\gamma_{\;\;,\beta)}T+ \gbar_{\gamma(\alpha} 
\LL^{\gamma\prime}_{\;\;,\beta)} T
+\LL^\delta_{\;\;,(\beta}\gbar_{\alpha)\gamma} \LL^\gamma_{\;\;,\delta}
+{1\over 2} \gbar_{\gamma\delta}\LL^\gamma_{\;\;,\alpha}\LL^\delta_{\;\;,\beta}
\nonumber \\
&&+\LL^\delta\left( {1\over 2} \gbar_{\alpha\beta,\gamma}
\LL^\gamma_{\;\;,\delta}
+\LL^\gamma_{\;\;,(\beta}\gbar_{\alpha)\gamma,\delta}+ {1\over 4} 
\gbar_{\alpha\beta,\gamma\delta} \LL^\gamma+\gbar_{\gamma(\alpha} 
\LL^\gamma_{\;\;,\beta)\delta} \right)~. \nonumber
\enar
It is possible to further decompose these metric perturbations into scalar,
vector, and tensor and to derive the transformation of the decomposed 
metric perturbations
by using Eqs.~(\ref{eq:decomp1}) and~(\ref{eq:decomp2}).
However, a few words in regard to spatial gauge-transformation are in order.
The spatial homogeneity of the background universe keeps the spatial
diffeomorphism intact to {\it all orders} in perturbations, 
and the physics is invariant under spatial 
gauge-transformation. However, it is well known \cite{BARDE80,MABE95,HWNO01}
that the perturbation variables
$(\beta,\gamma,\VBB_\alpha,\VCC_\alpha)$ change with the spatial transformation
$L$ or $L_\alpha$
at the linear order, carrying unphysical gauge modes,
\beeq
\tilde\beta=\beta-T+L'~,\quad\tilde\gamma=\gamma-L~,\quad
\tilde\VBB_\alpha=\VBB_\alpha+L_\alpha'~,\quad
\widetilde{\VCC}_\alpha=\VCC_\alpha-L_\alpha~.
\eneq
As physical quantities are invariant under spatial gauge-transformation, 
they can depend on these perturbation variables, only through
two combinations $\chi=a(\beta+\gamma')$ and 
$\pv_\alpha=\VBB_\alpha+\VCC'_\alpha$ that are invariant under spatial
gauge transformations \cite{BARDE88}. Writing the metric perturbations
in terms of these spatially invariant variables is readily achieved by
choosing a spatial gauge that leaves no unphysical gauge freedom 
$\LL^\alpha=0$ (i.e., $L=L^\alpha=0$).
We choose the $C$-gauge \cite{NOHW04} as our spatial gauge choice
\beeq
\tilde\gamma\equiv\gamma\equiv0~,\quad 
\widetilde{\VCC}_\alpha\equiv\VCC_\alpha\equiv0~,
\label{eq:Cgauge}
\eneq
which completely sets $L=0$, $L_\alpha=0$ to the linear order, while a choice
of temporal gauge is left free.
Combining the $C$-gauge choice
with any choice of a temporal gauge ($T=0$) at the linear order,
the second-order
gauge transformation of the spatial metric perturbation 
in Eq.~(\ref{eq:gtcc}) can be simplified as
\beeq
\widetilde{\CC}_{\alpha\beta}= \CC_{\alpha\beta}-\HH T^{(2)}\gbar_{\alpha\beta}
 - {1\over 2}\gbar_{\alpha\beta,\gamma} \LL^{(2)\gamma} -
{1\over2}\left(\gbar_{\gamma\alpha}\LL^{(2)\gamma}_{\;\;\;\;\;\;\;,\beta}   
+\gbar_{\gamma\beta}\LL^{(2)\gamma}_{\;\;\;\;\;\;\;,\alpha} \right)
=\CC_{\alpha\beta}-\HH T^{(2)}\gbar_{\alpha\beta}-\LL^{(2)}_{(\alpha|\beta)}~,
\eneq
and using Eqs.~(\ref{eq:decomp1}) and~(\ref{eq:decomp2})
its decomposed perturbations transform as
\beeq
\tilde\gamma^{(2)}=\gamma^{(2)}-L^{(2)}~,\quad
\tilde\VCC^{(2)}_\alpha=\VCC^{(2)}_\alpha-L^{(2)}_\alpha~.
\eneq
Therefore, the $C$-gauge condition in Eq.~(\ref{eq:Cgauge}) to the second order
in perturbations
completely removes the unphysical gauge freedom $\LL^\alpha=0$ to the same
order in perturbations, and we take the $C$-gauge as our spatial
gauge choice throughout the paper.
As opposed to choosing a (physical) temporal gauge, the
spatial gauge choice affects no physical quantities or the Einstein equations,
as the spatial diffeomorphism is unbroken symmetry.

With the spatial $C$-gauge choice but without any temporal gauge choice,
the metric perturbations transform as
\bear
\label{app:abc}
\tilde\alpha&=&\alpha - T^\prime -\HH T -\alpha^\prime T- 2\alpha T'
-2\alpha\HH T+{3\over2} T'T' + TT''+3\HH TT'+{1\over2}\HH'T^2+\HH^2T^2~,
\\
\tilde\BB_\alpha &=&\BB_\alpha- T_{,\alpha}- 2 \alpha T_{,\alpha}
-\left(\BB^\prime_\alpha+ 2 \HH\BB_\alpha\right)T-\BB_\alpha T'
 + 2T'T_{,\alpha} +T\left(T'_{,\alpha}+2\HH T_{,\alpha}\right)~,  
\\
\widetilde{\CC}_{\alpha\beta} &=&\CC_{\alpha\beta}-\HH T\gbar_{\alpha\beta}
+\BB_{(\alpha} T_{,\beta)} - \left(\CC^\prime_{\alpha\beta}
+ 2\HH\CC_{\alpha\beta} \right)T-{1\over 2} 
T_{,\alpha}T_{,\beta}+T\Bigg[\HH\gbar_{\alpha\beta}T'+{1\over2}
\left(2\HH^2+\HH'\right)\gbar_{\alpha\beta}T\Bigg]~,
\enar
and in terms of scalar, vector, and tensor components the metric is
\beeq
\AA = \alpha~, \quad 
\BB_\alpha = {1 \over a} \chi_{,\alpha}+\pv_\alpha~, \quad
\CC_{\alpha\beta} = \varphi~\gbar_{\alpha\beta}+\TCC_{\alpha\beta}~,
\eneq
where we used $\chi=a(\beta+\gamma')$ and 
$\pv_\alpha=\VBB_\alpha+\VCC_\alpha'$.
The above equations are fully general to the second order --- No temporal 
(physical) gauge choice is made, while unphysical spatial gauge freedom
is eliminated.
The gauge-transformation equations of the decomposed variables can be readily
derived by using Eqs.~(\ref{eq:decomp1}) and~(\ref{eq:decomp2}). 
In order to simplify the situation, we assume that the three-space
is flat ($K=0$). Therefore, the scalar perturbations transform as
\bear
\label{app:gtchi}
\tilde\chi&=&\chi- a T+ a TT'+a\HH T^2+a\Delta^{-1}\nabla^\alpha\left[
-2 \alpha T_{,\alpha}-{1 \over a}\left(\chi'+ \HH\chi\right)_{,\alpha}T
-(\pv'_\alpha+2\HH\pv_\alpha)T
-\left({1 \over a}\chi_{,\alpha}+\pv_\alpha\right)T'
+T'T_{,\alpha} \right]  \nonumber  \\
&&\hspace{-20pt}
-{a \over 2} \Delta^{-1} \left[\left({1\over a} \chi^{,\alpha}
+\pv^\alpha\right)T_{,\alpha}
-{1\over 2}T^{,\alpha}T_{,\alpha}-3\Delta^{-1}\nabla^\alpha\nabla^\beta\left(
{1 \over a} \chi_{,\alpha}T_{,\beta}-{1\over2}T_{,\alpha}T_{,\beta} 
+\pv_{(\alpha}T_{,\beta)}-(\TCC'_{\alpha\beta}+2\HH\TCC_{\alpha\beta})T
\right)\right]^\prime~, \\
\label{app:gtphi}
\tilde\varphi&=&\varphi-\HH T-\varphi'T-2\HH\varphi T+\HH TT'+{1\over 2}
\left(2\HH^2+\HH'\right)T^2+{1 \over 2}\left({1\over a}\chi^{,\alpha}
T_{,\alpha}+\pv_\alpha T^{,\alpha}
-{1\over2}T^{,\alpha}T_{,\alpha}\right)  \nonumber \\
&&
-{1\over2}\Delta^{-1}\nabla^\alpha\nabla^\beta\left[{1\over a}\chi_{,\alpha}
T_{,\beta} +\pv_{(\alpha}T_{,\beta)}-\left(\TCC'_{\alpha\beta}
+2\HH\TCC_{\alpha\beta}\right)T
-{1\over 2} T_{,\alpha}T_{,\beta}\right] ~,
\enar
and the vector and tensor perturbations transform as
\bear
\label{app:gtpv}
\tilde\pv_\alpha &=&\pv_\alpha
- 2 \alpha T_{,\alpha}
-{1\over a}\left(\chi^\prime_{,\alpha}+ \HH\chi_{,\alpha}\right)T
-\left(\pv^\prime_\alpha+ 2 \HH\pv_\alpha\right)T
-{\chi_{,\alpha}\over a} T'-\pv_\alpha T' - TT'_{,\alpha}\nonumber \\
&&
-\nabla_\alpha\Delta^{-1}\nabla^\beta\left[
- 2 \alpha T_{,\beta}
-{1\over a}\left(\chi^\prime_{,\beta}+ \HH\chi_{,\beta}\right)T
-\left(\pv^\prime_\beta+ 2 \HH\pv_\beta\right)T
-{\chi_{,\beta}\over a} T'-\pv_\beta T' - TT'_{,\beta}\right]\nonumber \\
&&
+2\Delta^{-1}\nabla^\beta\left[
{1\over a}\chi_{(,\alpha} T_{,\beta)} +\pv_{(\alpha} T_{,\beta)} 
- \left(\TCC^\prime_{\alpha\beta}
+ 2\HH\TCC_{\alpha\beta} \right)T-{1\over 2} T_{,\alpha}T_{,\beta}\right]'
\nonumber \\
&&
-2\Delta^{-1}\nabla_\alpha\Delta^{-1}\nabla^\beta\nabla^\gamma
\left[{1\over a}\chi_{(,\beta} T_{,\gamma)} +\pv_{(\beta} T_{,\gamma)} 
- \left(\TCC^\prime_{\beta\gamma}
+ 2\HH\TCC_{\beta\gamma} \right)T-{1\over 2} T_{,\beta}T_{,\gamma}
\right]'~, \\
\label{app:gttcc}
\widetilde\TCC_{\alpha\beta}&=&\TCC_{\alpha\beta}+
{1\over a}\chi_{(,\alpha} T_{,\beta)} +\pv_{(\alpha} T_{,\beta)} 
- \left(\TCC^\prime_{\alpha\beta}
+ 2\HH\TCC_{\alpha\beta} \right)T-{1\over 2} T_{,\alpha}T_{,\beta}
-{1\over3}\left[\left({1\over a}\chi^{,\gamma}+\pv^\gamma\right)T_{,\gamma}
-{1\over2}T^{,\gamma}T_{,\gamma}\right]\gbar_{\alpha\beta}\nonumber\\
&&\hspace{-20pt}
-{1\over2}\left(\nabla_\alpha\nabla_\beta-{1\over3}
\gbar_{\alpha\beta}\Delta\right)\Delta^{-1}
\bigg\{3\Delta^{-1}\nabla^\gamma\nabla^\delta\left[
{1\over a}\chi_{(,\gamma} T_{,\delta)} +\pv_{(\gamma} T_{,\delta)} 
- \left(\TCC^\prime_{\gamma\delta}
+ 2\HH\TCC_{\gamma\delta} \right)T-{1\over 2} T_{,\gamma}T_{,\delta}\right]
\nonumber \\
&&\hspace{-20pt}
-\left({1\over a}\chi^{,\gamma}+\pv^\gamma\right)T_{,\gamma}
+{1\over2}T^{,\gamma}T_{,\gamma}\bigg\}
-2\nabla_{(\alpha}\Delta^{-1}\bigg\{\nabla^\gamma\left[
{1\over 2a}\left(\chi_{,\beta)} T_{,\gamma} +\chi_{,\gamma}T_{,\beta)}\right)
+\pv_{\beta)} T_{,\gamma} - \left(\TCC^\prime_{\beta)\gamma}
+ 2\HH\TCC_{\beta)\gamma} \right)T-{1\over 2} T_{,\beta)}T_{,\gamma}\right]
\nonumber \\
&&\hspace{-20pt}
-\nabla_{\beta)}\Delta^{-1}\nabla^\gamma\nabla^\delta
\left[{1\over a}\chi_{(,\gamma} T_{,\delta)} +\pv_{(\gamma} T_{,\delta)} 
- \left(\TCC^\prime_{\gamma\delta}
+ 2\HH\TCC_{\gamma\delta} \right)T-{1\over 2} T_{,\gamma}T_{,\delta}\right]
\bigg\}~.
\enar
It is evident that scalar, vector, and tensor components 
mix together due to the nonlinear quadratic terms, present in the second-order 
gauge transformation. Furthermore, even with no vector or tensor at the 
linear order, the second-order scalar perturbations generate the second-order
vector and tensor perturbations. Nevertheless, the equations greatly
simplify in the absence of linear order vector and tensor perturbations,
especially for scalar perturbations.

Equations~(\ref{app:gtchi})$-$(\ref{app:gttcc}) are completely general to
the second order in perturbations, and no physical gauge choice is made yet.
Furthermore, it is apparent  that combined with the spatial $C$-gauge,
a proper temporal
 gauge choice at the linear order provides a valid gauge choice
at the second order, e.g., if we choose a gauge condition at the linear order
\beeq
\tilde\chi^{(1)}=\chi^{(1)}=0~,\quad T^{(1)}=0~,
\eneq
the transformation equation in Eq.~(\ref{app:gtchi}) 
at the second order in perturbations takes the form
\beeq
\tilde\chi^{(2)}=\chi^{(2)}-aT^{(2)}~,
\eneq
identical to its linear order transformation equation. Therefore, by 
choosing a gauge condition to the second order, $\chi^{(1,2)}=0$ in this
example, we completely fix the gauge condition to the second order in 
perturbations, leaving no gauge ambiguities.

Similarly to the transformation in metric tensor, 
the coordinate transformation in Eq.~(\ref{eq:gt}) induces
the vector transformation
\beeq
\tilde V^a(\hat x^e)={\partial \tilde x^a\over \partial x^b}~V^b(x^e)~,
\eneq
and by evaluating the transformed vector at the same coordinate $x^e$,
we can derive the vector gauge-transformation relation as
\beeq
\label{app:etc}
\widetilde{\VV}^\alpha=\VV^\alpha+\HH T\VV^\alpha-T\VV^{\alpha\prime}~,
\eneq
and the perturbations in photon wavevector transform as
\bear
\widetilde{\dnu}&=&\dnu+{d\over d\cc}T+2\HH T(1+\dnu)-\dnu' T
-TT^{\prime\prime}- \left(\HH'-2\HH^2 \right)T^2
+TT^{\prime}_{,\alpha} e^\alpha-2\HH TT_{,\alpha} e^\alpha~,\\
\widetilde{\dea}^\alpha &=&\dea^\alpha+ 2\HH T\left(e^\alpha+\dea^\alpha\right)
- T \dea^{\alpha\prime}
-2\HH TT^{\prime}  e^\alpha-\left( \HH'-2\HH^2 \right) T^2e^\alpha ~.
\enar

\subsection{Second-order gauge-invariant variables}
\label{app:gi}
Here we construct second-order gauge-invariant variables. First, 
comparing to the linear-order calculations, we 
discuss the difference in the second-order calculation by explicitly
constructing a second-order gauge-invariant variable. We then give expressions
for other second-order gauge-invariant variables used in the text.
Our construction of second-order gauge-invariant variables follows the work
in \cite{NOHW04}, but without the restriction that there is no vector or 
tensor at the linear order.

As an example, we construct the second-order gauge-invariant variable~$\px$.
The linear-order gauge-transformation equations for~$\chi$ and~$\varphi$ 
in Eqs.~(\ref{app:gtchi}) and~(\ref{app:gtphi}) are   
\beeq
\tilde\chi^{(1)}=\chi^{(1)}-aT^{(1)}~,\quad
\tilde\varphi^{(1)}=\varphi^{(1)}-\HH~T^{(1)}~,
\eneq
and we can easily construct a linear-order gauge-invariant variable
\beeq
\px^{(1)}\equiv\varphi^{(1)}-H\chi^{(1)}~.
\eneq
The notation of the gauge-invariant variable is set up such that~$\px$ becomes
$\varphi$ when the gauge condition $\chi=0$ is adopted
(similarly, we can also construct a gauge-invariant variable
$\chi_\varphi\equiv-\px/H=\chi-\varphi/H$, such that $\chi_\varphi$ becomes
$\chi$ when the gauge condition $\varphi=0$ is adopted).
Therefore, it is desirable to construct such gauge-invariant variables
at the second order. The simplest guess is to extend the definition of~$\px$
at the linear order to the second order. 
Using Eqs.~(\ref{app:gtchi}) and~(\ref{app:gtphi}), we verify that the
simplest choice transforms as
\bear
\label{app:gtpx}
\tilde\varphi-H\tilde\chi
&=&\varphi-H\chi-\varphi'T-2\HH\varphi T+{1\over 2}\HH'T^2
+{1 \over 2}\left({1\over a}\chi^{,\alpha}
T_{,\alpha}+\pv^\alpha T_{,\alpha}-{1\over2}T^{,\alpha}T_{,\alpha}\right)  \\
&&-{1\over2}\Delta^{-1}\nabla^\alpha\nabla^\beta\left[
\left({1\over a}\chi_{,\alpha}+\pv_\alpha\right)T_{,\beta} 
-(\TCC'_{\alpha\beta}+2\HH\TCC_{\alpha\beta})T
-{1\over 2} T_{,\alpha}T_{,\beta}\right] \nonumber \\
&&
-\HH\Delta^{-1}\nabla^\alpha\left[
-2 \alpha T_{,\alpha}-{1 \over a}\left(\chi'+ \HH\chi\right)_{,\alpha}T
-(\pv'_\alpha+2\HH\pv_\alpha)T-\left({1 \over a}\chi_{,\alpha}
+\pv_\alpha\right)T'+T'T_{,\alpha} \right]    \nonumber\\
&&
+{\HH \over 2} \Delta^{-1} \left[\left({1\over a} \chi^{,\alpha}
+\pv^\alpha\right)T_{,\alpha}
-{1\over 2}T^{,\alpha}T_{,\alpha}-3\Delta^{-1}\nabla^\alpha\nabla^\beta\left(
{1 \over a} \chi_{,\alpha}T_{,\beta}-{1\over2}T_{,\alpha}T_{,\beta} 
+\pv_{(\alpha}T_{,\beta)}-(\TCC'_{\alpha\beta}+2\HH\TCC_{\alpha\beta})T
\right)\right]^\prime~. \nonumber
\enar
It is evident that this particular combination
is {\it not} gauge-invariant to the second-order. However, the transformation
in Eq.~(\ref{app:gtpx}) suggests that a quadratic correction to the simplest
combination be needed to construct a second-order gauge-invariant variable
and the correction should vanish if we choose the gauge condition $\chi=0$.
Since the quadratic correction only involves the linear-order transformation,
we use
\beeq
T^{(1)}={1\over a}\left(\chi^{(1)}-\tilde\chi^{(1)}\right)~,
\label{app:linear}
\eneq
to find the quadratic correction for the gauge-invariant variable that
vanishes if we choose the gauge-condition $\chi=0$.

By substituting Eq.~(\ref{app:linear}) into Eqs.~(\ref{app:gtchi}) 
and~(\ref{app:gtphi}), we have
\bear
\label{app:expli}
\tilde\chi&=&\chi- a T+ 
H(\chi\tilde\chi-\chi^2)-(\chi-\tilde\chi)\dot{\tilde\chi} 
-\Delta^{-1}\nabla^\alpha\bigg[2\ax(\chi-\tilde\chi)_{,\alpha}
+\left(\dot\chi-H\chi\right)\chi_{,\alpha}
-\left(\dot{\tilde\chi}-H\tilde\chi\right)\tilde\chi_{,\alpha} \nonumber \\
&&
+(\pv'_\alpha+2\HH\pv_\alpha)(\chi-\tilde\chi)+\pv_\alpha(\chi-\tilde\chi)'
\bigg]
-a^2\Delta^{-1} \bigg[{1\over 4a^2}\left(\chi^{,\alpha}\chi_{,\alpha}
-\tilde\chi^{,\alpha}\tilde\chi_{,\alpha}\right)
+{1\over2a}\pv^\alpha(\chi-\tilde\chi)_{,\alpha} \nonumber \\
&&
-{3\over 2a}\Delta^{-1}\nabla^\alpha\nabla^\beta\bigg(
{\chi_{,\alpha}\chi_{,\beta}-\tilde\chi_{,\alpha}\tilde\chi_{,\beta}\over2a}
+\pv_\alpha(\chi-\tilde\chi)_{,\beta}-(\TCC'_{\alpha\beta}
+2\HH\TCC_{\alpha\beta})(\chi-\tilde\chi)\bigg)\bigg]^\cdot~, \\
\tilde\varphi&=&
\varphi-\HH T-(\dot\px+2H\px)(\chi-\tilde\chi)-{1\over2}(H^2+\dot H)
(\chi^2-\tilde\chi^2)+H^2(\chi\tilde\chi-\chi^2)
-H(\chi-\tilde\chi)\dot{\tilde\chi} 
+{1\over4a^2}\left(\chi^{,\alpha}\chi_{,\alpha}
-\tilde\chi^{,\alpha}\tilde\chi_{,\alpha}\right)\nonumber \\
&&
+{1\over2a}\pv^\alpha(\chi-\tilde\chi)_{,\alpha}-
{1\over2a}\Delta^{-1}\nabla^\alpha\nabla^\beta
\bigg[{\chi_{,\alpha}\chi_{,\beta}-\tilde\chi_{,\alpha}
\tilde\chi_{,\beta}\over2a}
+\pv_{\alpha}(\chi-\tilde\chi)_{,\beta}-(\TCC'_{\alpha\beta}
+2\HH\TCC_{\alpha\beta})(\chi-\tilde\chi)\bigg] ~.
\enar
Collecting terms with tilde on the left-hand side, 
we obtain the second-order gauge-invariant variable
\bear
\label{app:pxgi}
\varphi_\chi   &\equiv&  \varphi - H \chi
       - \left( \dot \varphi_\chi + 2 H \varphi_\chi \right) \chi
       - {1 \over 2} \left( \dot H + H^2 \right) \chi^2 
+ H \Delta^{-1} \nabla^\alpha \bigg[ 2 \alpha_\chi \chi_{,\alpha}
 + \left( \dot \chi - H \chi \right) \chi_{,\alpha} 
+(\pv'_\alpha+2\HH\pv_\alpha)\chi+\pv_\alpha\chi'\bigg]\nonumber \\
&&
+ {1\over4a^2}~\chi^{,\alpha} \chi_{,\alpha}
+{1\over2a}\pv^\alpha\chi_{,\alpha} - {1\over2 a}
\Delta^{-1} \nabla^\alpha \nabla^\beta
 \bigg[ {1\over2a}\chi_{,\alpha} \chi_{,\beta}+\pv_\alpha\chi_{,\beta}
-(\TCC'_{\alpha\beta}+2\HH\TCC_{\alpha\beta})\chi\bigg]   \nonumber \\
&&     
+ a^2 H \Delta^{-1} \left[
 {1 \over 4a^2} \chi^{,\alpha} \chi_{,\alpha}+{1\over2a}\pv^\alpha
\chi_{,\alpha}
-  {3 \over 2a} \Delta^{-1} \nabla^\alpha \nabla^\beta
\bigg( {1\over2a}\chi_{,\alpha} \chi_{,\beta} 
+\pv_\alpha\chi_{,\beta}-(\TCC'_{\alpha\beta}+2\HH\TCC_{\alpha\beta})\chi
\bigg) \right]^\cdot~, 
\enar
and we check its gauge-invariance by explicitly performing transformation.
The gauge-invariant variable~$\px$ has the property that $\px\RA\varphi$
under the gauge condition $\chi=0$.
Similar calculations can be performed to construct the 
second-order gauge-invariant variable
\bear
\label{app:etcgi}
\ax&\equiv&\alpha - \dot\chi-\dot\ax\chi-2\ax\dot\chi+{1\over2}\dot H\chi^2
+H\chi\dot\chi-{1\over2}\dot\chi^2
+\Delta^{-1}\nabla^\alpha\bigg[2\ax\chi_{,\alpha}
+\left(\dot\chi-H\chi\right)\chi_{,\alpha}
+(\pv'_\alpha+2\HH\pv_\alpha)\chi+\pv_\alpha\chi'\bigg]^\cdot\nonumber \\
&&\hspace{-50pt}
+\Bigg[a^2\Delta^{-1} \bigg[{1\over 4a^2}\chi^{,\alpha}\chi_{,\alpha}
+{1\over2a}\pv^\alpha\chi_{,\alpha} 
-{3\over 2a}\Delta^{-1}\nabla^\alpha\nabla^\beta\bigg(
{1\over2a}\chi_{,\alpha}\chi_{,\beta}
+\pv_\alpha\chi_{,\beta}-(\TCC'_{\alpha\beta}
+2\HH\TCC_{\alpha\beta})\chi\bigg)\bigg]^\cdot\Bigg]^\cdot
\equiv\alpha-\dot\chi+\ax^\qqq~,~~~
\enar
where we defined the quadratic correction terms that vanish under the 
gauge condition indicated in the subscript (similarly, we also define
$\px\equiv\varphi-H\chi+\px^\qqq$ in Eq.~[\ref{app:pxgi}]).

We continue to repeat the exercise to construct gauge-invariant variables
for four vectors. Perturbations to the photon wavevector and four velocity
can be rearranged to define second-order gauge-invariant variables as
\bear
\label{app:deag}
\deag&\equiv&\dea^\alpha+ 2H\chi\left(e^\alpha+\deag\right)-\chi\dot\deag
+(\dot H-3H^2)\chi^2e^\alpha \nonumber \\
&&
-2He^\alpha\Delta^{-1}\nabla^\beta\bigg[2\ax\chi_{,\beta}
+\left(\dot\chi-H\chi\right)\chi_{,\beta}
+(\pv'_\beta+2\HH\pv_\beta)\chi+\pv_\beta\chi'\bigg]  \nonumber \\
&&
-2a^2He^\alpha
\Delta^{-1} \bigg[{1\over 4a^2}\chi^{,\beta}\chi_{,\beta}
+{1\over2a}\pv^\beta\chi_{,\beta}
-{3\over 2a}\Delta^{-1}\nabla^\beta\nabla^\gamma\bigg(
{\chi_{,\beta}\chi_{,\gamma}\over2a}
+\pv_\beta\chi_{,\gamma}-(\TCC'_{\beta\gamma}
+2\HH\TCC_{\beta\gamma})\chi\bigg)\bigg]^\cdot~,\\
\label{app:dnug}
\dnug&\equiv&\dnu+H\chi+\dot\chi-{1\over a}\chi_{,\alpha}e^\alpha
+2H\chi\dnug-\chi~\dot\dnug+\dnug(\dot\chi-H\chi)-{1\over a} \chi_{,\alpha}
\deag-\dot\chi^2-H^2\chi^2-2H\chi\dot\chi\nonumber\\
&&
+{3H\over a}\chi\chi_{,\alpha}e^\alpha
+{1\over a}\dot\chi\chi_{,\alpha}e^\alpha 
-H\Delta^{-1}\nabla^\alpha\bigg[2\ax\chi_{,\alpha}
+\left(\dot\chi-H\chi\right)\chi_{,\alpha}
+(\pv'_\alpha+2\HH\pv_\alpha)\chi+\pv_\alpha\chi'\bigg]\nonumber \\
&&
-a^2H\Delta^{-1} \bigg[{1\over 4a^2}\chi^{,\alpha}\chi_{,\alpha}
+{1\over2a}\pv^\alpha\chi_{,\alpha} 
-{3\over 2a}\Delta^{-1}\nabla^\alpha\nabla^\beta\bigg(
{\chi_{,\alpha}\chi_{,\beta}\over2a}
+\pv_\alpha\chi_{,\beta}-(\TCC'_{\alpha\beta}
+2\HH\TCC_{\alpha\beta})\chi\bigg)\bigg]^\cdot \nonumber \\
&&-{1\over a}{d\over d\cc}\bigg[
\Delta^{-1}\nabla^\alpha\bigg(2\ax\chi_{,\alpha}
+\left(\dot\chi-H\chi\right)\chi_{,\alpha}
+(\pv'_\alpha+2\HH\pv_\alpha)\chi+\pv_\alpha\chi'\bigg)\bigg]\nonumber \\
&&-{1\over a}{d\over d\cc}\bigg\{
a^2\Delta^{-1} \bigg[{1\over 4a^2}\chi^{,\alpha}\chi_{,\alpha}
+{1\over2a}\pv^\alpha\chi_{,\alpha} 
-{3\over 2a}\Delta^{-1}\nabla^\alpha\nabla^\beta\bigg(
{\chi_{,\alpha}\chi_{,\beta}\over2a}
+\pv_\alpha\chi_{,\beta}-(\TCC'_{\alpha\beta}
+2\HH\TCC_{\alpha\beta})\chi\bigg)\bigg]^\cdot\bigg\}~,~~~\\
\label{app:VV}
\VV^\alpha_\chi&\equiv&\VV^\alpha+H\chi~\VV^\alpha-\chi~\dot\VV^\alpha~,
\enar
where the spatial part of the four velocity itself is gauge-invariant at
the linear order. Finally,
gauge-invariant variables for vector and tensor in metric perturbations
are
\bear
\label{app:pv}
\pv_{\chi\alpha} &\equiv&\pv_\alpha-  {2\over a}\ax\chi_{,\alpha}
+{1\over a}(\chi\dot\chi_{,\alpha}-H\chi\chi_{,\alpha}) 
-\left(\dot\pv_\alpha+ 2 H\pv_\alpha\right)\chi-\pv_\alpha (\dot\chi-H\chi)
\nonumber \\
&&
-\nabla_\alpha\Delta^{-1}\nabla^\beta\left[- {2\over a}\ax\chi_{,\beta}
+{1\over a}(\chi\dot\chi_{,\alpha}-H\chi\chi_{,\alpha})
-\left(\dot\pv_\beta+ 2 H\pv_\beta\right)\chi-\pv_\beta(\dot\chi-H\chi)
\right]\nonumber \\
&&
+2a\Delta^{-1}\nabla^\beta\left[{1\over2a^2}\chi_{,\alpha}\chi_{,\beta}
+{1\over a}\pv_{(\alpha}\chi_{,\beta)}- \left(\dot\TCC_{\alpha\beta}
+ 2H\TCC_{\alpha\beta} \right)\chi\right]^\cdot
\nonumber \\
&&
-2a\Delta^{-1}\nabla_\alpha\Delta^{-1}\nabla^\beta\nabla^\gamma
\left[{1\over2a^2}\chi_{,\beta}\chi_{,\gamma}+{1\over a}\pv_{(\alpha}
\chi_{,\beta)}- \left(\dot\TCC_{\beta\gamma}
+ 2H\TCC_{\beta\gamma} \right)\chi\right]^\cdot~, \\
\label{app:tcc}
\TCC_{\chi\alpha\beta}&\equiv&\TCC_{\alpha\beta}+
{1\over2a^2}\chi_{,\alpha}\chi_{,\beta}
+{1\over a}\pv_{(\alpha} \chi_{,\beta)} - \left(\dot\TCC_{\alpha\beta}
+ 2H\TCC_{\alpha\beta} \right)\chi-{1\over3a}~\gbar_{\alpha\beta}
\left({1\over 2a}\chi^{,\gamma}+\pv^\gamma\right)\chi_{,\gamma}
\nonumber\\
&&\hspace{-40pt}
-{1\over2}\left(\nabla_\alpha\nabla_\beta-{1\over3}
\gbar_{\alpha\beta}\Delta\right)\Delta^{-1}
\left\{3\Delta^{-1}\nabla^\gamma\nabla^\delta\left[
{1\over 2a^2}\chi_{,\gamma}\chi_{,\delta} 
+{1\over a}\pv_{(\gamma}\chi_{,\delta)} - \left(\dot\TCC_{\gamma\delta}
+ 2H\TCC_{\gamma\delta} \right)\chi\right]
-{1\over a}\left({1\over 2a}\chi^{,\gamma}+\pv^\gamma\right)\chi_{,\gamma}
\right\} \nonumber\\
&&
-2\nabla_{(\alpha}\Delta^{-1}\bigg\{\nabla^\gamma\left[
{1\over 2a^2}\chi_{,\beta)} \chi_{,\gamma} 
+{1\over a}\pv_{\beta)}\chi_{,\gamma} - \left(\dot\TCC_{\beta)\gamma}
+ 2H\TCC_{\beta)\gamma} \right)\chi\right]
\nonumber \\
&&
-\nabla_{\beta)}\Delta^{-1}\nabla^\gamma
\nabla^\delta
\left[{1\over 2a^2}\chi_{,\gamma} \chi_{,\delta} 
+{1\over a}\pv_{(\gamma} \chi_{,\delta)} - \left(\dot\TCC_{\gamma\delta}
+ 2H\TCC_{\gamma\delta} \right)\chi\right]\bigg\}~. 
\enar
It is obvious that the second-order gauge-invariant variables are more
complicated than its linear-order piece, due to the mixing of
scalar, vector, and tensor contributions. Exactly for this reason,
the second-order gauge-invariant variable for tensor~$\TCC_{\alpha\beta}$
(and also vector $\pv_{\alpha}$) requires a choice of scalar gauge condition.
It is noted that other choices of scalar gauge condition can be made
to construct these second-order gauge-invariant variables, and even with
no vector or tensor at the linear order one needs a choice of scalar
gauge condition for second-order vector and tensor gauge-invariant variables,
as scalar contributions generate vector and tensor
to the second order in perturbations.

\section{Photon Wavevector}
\label{app:photon}
In this section, we explicitly construct second-order tetrads 
to derive the photon wavevector $k^a$ in the FRW frame in terms of local
observable quantities. We then clarify the relation of the photon wavevector
to the observed angles $(\ttt,\pp)$ in the presence
of additional extra degree of freedom supplied by the conformal transformation
of the metric. The basic description of the geometric optics can be found
in \cite{MTW,SACHS61,KRSA66}.

\subsection{Second-order tetrads in the observer rest frame}
\label{app:tetrads}
Here we construct a second-order orthonormal basis in the observer rest
frame defined by the observer's four velocity~$u^a$. The time-like 
velocity ($-1=u_au^a$) of the observer defines the (proper)
time-direction $[e_t]^a\equiv u^a$ 
in the local Lorentz frame and its hypersurface orthogonal 
to~$u^a$. Three spacelike four vectors $[e_i]^a$ ($1=[e_i]^a[e_i]_a$,
$i=x,y,z$)
can be further defined to serve as spatial directions in the observer rest
frame. These four orthonormal vectors are called tetrads.
In the local Lorentz frame, the metric is Minkowsky 
($g_{\mu\nu}^L=\eta_{\mu\nu}=g_{ab}[e_\mu]^a[e_\nu]^b$, $\mu,\nu=t,x,y,z$), 
and the tetrads are simply unit vectors:
$[e_t]=(1,0,0,0)$ and $[e_i]=(0,\delta_i^\alpha)$. However, we are interested
in tetrad expressions in an inhomogeneous expanding (FRW) universe with the
metric described by Eqs.~(\ref{eq:bgmetric}) and~(\ref{eq:abc}).

Accounting for the expansion, the tetrads in a homogeneous universe are
\beeq
\label{eq:tet0}
[e_t]^a=u^a={1\over a}(1,0,0,0)~,~~~[e_x]^a={1\over a}(0,1,0,0)~,
~~~[e_y]^a={1\over a}(0,0,1,0)~,~~~[e_z]^a={1\over a}(0,0,0,1)~,
\eneq
with the tetrad index~$\mu$ of $[e_\mu]^a$
raised or lowered by~$\eta_{\mu\nu}$, while
the FRW index~$a$ is raised or lowered by $g_{ab}$.\footnote{In the local
Lorentz frame of the observer, the time coordinate $x_L^t$ is equivalent to the
proper time of the observer. Therefore, the path $x^a_F$ of the observer
in a FRW coordinate
parametrized by the proper time yields the observer four velocity
$u^a=\partial x^a_F/\partial x^t_L=[e_t]^a$. While our convention for tetrads
is set consistent with this notion ($[e_t]^a\equiv u^a=-[e^t]^a$),
it can be set with different sign,
$[e^t]^a\equiv u^a=-[e_t]^a$, which simply changes the
direction of proper time backward, leaving the construction of other
spatial directions $[e_i]^a$ unaffected.}
In an inhomogeneous universe, we add perturbations to describe the deviation
from homogeneity, defining the observer four velocity in an inhomogeneous
universe:
\beeq
[e_t]^a=u^a\equiv{1\over a}\left(1+\dUU^0,~\VV^\alpha\right)~.
\eneq
With the normalization condition $-1=g_{ab}[e_t]^a[e_t]^b$,
we have the observer four velocity to the second order in perturbations
\beeq
[e_t]^a={1\over a}\left[1-\AA+{3\over2}\AA^2
+{1\over2}\VV^\beta\VV_\beta-\VV^\beta\BB_\beta~,~\VV^\alpha\right]~,\quad
[e_t]_a=a\left[-1-\AA+{1\over2}\AA^2-{1\over2}\VV^\beta\VV_\beta
~,~\VV_\alpha-\BB_\alpha+\AA\BB_\alpha+2\VV^\beta\CC_{\alpha\beta}\right]~,
\eneq
consistent with Eq.~(\ref{eq:fourv}). The remaining spatial directions are
also derived by using the orthonormality condition
$\delta_{ij}=g_{ab}[e_i]^a[e_j]^b$ and $0=g_{ab}[e_t]^a[e_i]^b$ as
\bear
[e_i]^a&=&{1\over a}\left[\VV_i-\BB_i+2\AA\BB_i-\AA\VV_i+\CC_{i\beta}
\left(\VV^\beta+\BB^\beta\right)~,
~\delta^\alpha_i-\CC^\alpha_i+{1\over2}\left(\VV_i\VV^\alpha-\BB_i\BB^\alpha
\right)+{3\over2}\CC^\beta_i\CC^\alpha_\beta\right]~,\\
\label{eq:tet2}
{[e_i]}_a&=&a\left[-\VV_i-\AA\VV_i-\CC_{i\beta}\VV^\beta~,
~\delta_{i\alpha}+\CC_{i\alpha}+{1\over2}\left(\BB_i\BB_\alpha-\CC_{i\beta}
\CC^\beta_\alpha+\VV_i\VV_\alpha\right)-\VV_i\BB_\alpha\right]~.
\enar
To ensure that our second-order construction of tetrads is correct, we
reconstruct the FRW metric $g_{ab}$ by transforming the local 
coordinate $x^\mu_L$ to the FRW coordinate $x^a_F$ 
\beeq
g_{ab}(x_F)={\partial x_L^\mu\over\partial x_F^a}~
{\partial x_L^\nu\over\partial x_F^b}~\eta_{\mu\nu}(x_L)
=[e^\mu]_a[e^\nu]_b~\eta_{\mu\nu}=-[e^t]_a[e^t]_b
+\sum_i[e^i]_a[e^i]_b~,
\eneq
and check if each component of the reconstructed metric is identical
to the metric defined in Eqs.~(\ref{eq:bgmetric}) and~(\ref{eq:abc}).

\subsection{Photon wavevector in FRW coordinates}
\label{app:wavefrw}
The photon propagation direction is the direction orthogonal to the 
hypersurface defined by the same phase $\theta=\bdi{k}\cdot\bdi{x}_L-\omega t$.
The components of the photon wavevector in the local Lorentz frame are
\beeq
k^\mu_L=\eta^{\mu\nu}\theta_{,\nu}=\left(\omega~,~\bdi{k}\right)
=2\pi\nu~(1~,~-\nhat)~,
\label{eq:lwave}
\eneq
where the angular frequency is $\omega=2\pi\nu$, $k_L=|\bdi{k}|=2\pi/\lambda$,
the speed of light
$c=\lambda\nu=1$, and we put the subscript~$L$ to emphasize that the components
are written in the local Lorentz frame (as opposed to the FRW frame).
We defined a unit directional vector $\nhat$
for photon propagation measured by the observer, $\nhat\propto-\bdi{k}$.
The photon frequency measured by the observer is then
\beeq
-\eta_{\mu\nu}u^\mu_L k^\nu_L=\omega=2\pi\nu~,
\eneq
where $u^\mu_L=[e_t]=(1,0,0,0)$ in the local Lorentz frame.
Since the photon wavevector is expressed in terms of physical quantities 
(the observed frequency and angle),
there are no additional degrees of freedom associated with the photon
wavevector in Eq.~(\ref{eq:lwave}). 

Now we compute the photon wavevector in a FRW coordinate by transforming the
photon wavevector in Eq.~(\ref{eq:lwave}) as
\bear
\label{eq:fwave}
k^a&=&{\partial x_F^a\over\partial x_L^\mu}~k^\mu_L=[e_\mu]^ak^\mu_L~,\\
\label{eq:fwavet}
k^0&=&{2\pi\nu\over a}\left\{1-\AA-n^i(\VV_i-\BB_i)
+{3\over2}\AA^2+{1\over2}\VV^\beta\VV_\beta-\VV^\beta\BB_\beta
-n^i\left[2\AA\BB_i-\AA\VV_i+\CC_{i\beta}\left(\VV^\beta+
\BB^\beta\right)\right]\right\}~,\\
\label{eq:fwaves}
k^\alpha&=&{2\pi\nu\over a}\left\{-n^\alpha+\VV^\alpha+n^i\CC_i^\alpha
-n^i\left[{1\over2}\left(\VV_i\VV^\alpha-\BB_i\BB^\alpha
\right)+{3\over2}\CC^\beta_i\CC^\alpha_\beta\right]\right\}~,
\enar
where $n^i$ is the spatial component of the unit directional vector $\nhat$
in a local Lorentz frame, other perturbation quantities are those
in a FRW frame, and
the repeated indices indicate the summation over the spatial 
components.\footnote{Greek indices $\alpha,\beta$ are used to represent 
the spatial components of four vectors in a FRW coordinate, and Latin indicies
$i,j$ are used to represent those in a Local Lorentz frame. However, as far
as the summation is concerned, there is no distinction, as the three vectors
in a FRW coordinate are based on the mean three-metric $\gbar_{\alpha\beta}$
in a flat universe ($K=0$).
Nevertheless, it is noted that those three vectors have different values of
their components depending on frames.} Because of the observer velocity
$\VV^\alpha$ and the gravitational potential~$\AA$, the components of the
photon wavevector appear different in a FRW coordinate, but its physical
interpretation depends on gauge choice. The spatial photon direction $k^\alpha$
in a FRW coordinate is different from that $n^\alpha$
in the observer rest frame, because the observer is not at rest in the FRW
frame. However, the photon 
frequency measured by the observer is a Lorentz scalar, independent of frame:
\beeq
-g_{ab}u^ak^b=-\eta_{\mu\nu}u_L^\mu k^\nu_L=\omega=2\pi\nu~,
\eneq
which sets the affine parameter~$\oo$ as in Eq.~(\ref{eq:affineoo})
--- Given the spacetime metric $g_{ab}$ and locally measured observables
$(\nu,~\nhat)$,
the wavevector $k^a$ in Eq.~(\ref{eq:fwave})
is therefore completely set in terms of physical quantities.

\subsection{Normalization constant}
\label{app:nor}
With the conformal transformation relation in Eq.~(\ref{eq:conf})
and the wavevector in Eqs.~(\ref{eq:fwavet}) and~(\ref{eq:fwaves}),
we have one degree of freedom in the overall amplitude
of the photon wavevector, given the metric tensor~$g_{ab}$:
\beeq
\label{eq:cpf}
\CK^a=\NC a^2k^a\propto\NC\nu a~.
\eneq
While the normalization
coefficient~$\NC$ is constant, the photon frequency measured by local observers
changes at each spacetime due to redshift and perturbations, and
so does the product $\NC\nu a$.
The conformally transformed photon wavevector
can be explicitly written as
\bear
\label{eq:cpf0}
\CK^0&=&2\pi\NC \nu a\left\{1-\AA-n^i(\VV_i-\BB_i)
+{3\over2}\AA^2+{1\over2}\VV^\beta\VV_\beta-\VV^\beta\BB_\beta
-n^i\left[2\AA\BB_i-\AA\VV_i+\CC_{i\beta}\left(\VV^\beta+
\BB^\beta\right)\right]\right\}~,\\
\label{eq:cpfa}
\CK^\alpha&=&2\pi\NC \nu a\left\{-n^\alpha+\VV^\alpha+n^i\CC_i^\alpha
-n^i\left[{1\over2}\left(\VV_i\VV^\alpha-\BB_i\BB^\alpha
\right)+{3\over2}\CC^\beta_i\CC^\alpha_\beta\right]\right\}~.
\enar
Noting that due to expansion of the universe
the photon frequency redshifts as $\bar\nu\propto1/a$ 
in a homogeneous universe, we have the background relation and the 
normalization constant as
\beeq
\CK^a=\bar\NC a^2 k^a=2\pi\bar\NC\bar\nu a(1~,-n^\alpha)
\equiv(1~,-e^\alpha)~,\quad
\bar\NC={1\over2\pi\bar\nu a}={1\over2\pi\bar\nu_0}~,\quad 
e^\alpha=n^\alpha~,
\label{eq:bgcon}
\eneq
where $\bar\nu\equiv\bar\nu_0/a$ and $\bar\nu_0$ is a constant.
Equation~(\ref{eq:bgcon}) indicates that all the 
observers along the photon path measures the same direction $n^\alpha=e^\alpha$
(constant) and infers the same frequency $\nu$ based on the observed redshift
in a homogeneous universe.

Therefore, our parametrization of the photon wavevector in Eq.~(\ref{eq:wave})
is completely general with perturbations described by $(\dnu,~\dea^\alpha)$,
subject to the null condition in Eq.~(\ref{eq:null})
and the geodesic equations~(\ref{eq:tempo}) and~(\ref{eq:spatial}).
However, these perturbations are related to the physical quantities
$(\nu,~\nhat)$
measured by local observers at each spacetime point with only one degree
of freedom in the overall amplitude~$\NC$, arising from 
the conformal transformation. 
This can be further understood as follows. Splitting the photon frequency 
and the normalization constant into the mean
and its fluctuation as
\beeq
\label{eq:NCsplit}
\nu\equiv{\bar\nu_0\over a}(1+\dpnu)~,\qquad \NC\equiv\bar\NC(1+\DNC)~.
\eneq
The perturbations of the photon wavevector are related to the normalization
constant as
\bear
\label{eq:dnuC1}
\dnu^{(1)}&=&-\AA-n^\alpha(\VV_\alpha-\BB_\alpha)+\DNC+\dpnu~,\\
\label{eq:deaC1}
\dea^{\alpha(1)}&=&-\VV^\alpha-n^\beta\CC_\beta^\alpha
+n^\alpha(\DNC+\dpnu)~,\\
\label{eq:dnuC2}
\dnu^{(2)}&=&-\AA-n^\beta(\VV_\beta-\BB_\beta)
+{3\over2}\AA^2+{1\over2}\VV^\beta\VV_\beta-\VV^\beta\BB_\beta
-n^\beta\left[2\AA\BB_\beta-\AA\VV_\beta+\CC_{\beta\gamma}\left(\VV^\gamma+
\BB^\gamma\right)\right] \nonumber \\
&&+(\DNC+\dpnu)\left[1-\AA-n^\beta(\VV_\beta-\BB_\beta)\right]+\DNC~\dpnu~,\\
\label{eq:deaC2}
\dea^{\alpha(2)}&=&-\VV^\alpha-n^\beta\CC_\beta^\alpha
+n^\gamma\left[{1\over2}\left(\VV_\gamma\VV^\alpha
-\BB_\gamma\BB^\alpha\right)+{3\over2}\CC^\beta_\gamma\CC^\alpha_\beta\right]
\nonumber \\
&&+(\DNC+\dpnu)\left(n^\alpha-\VV^\alpha-\CC^\alpha_\beta n^\beta\right)
+\DNC~\dpnu ~n^\alpha~,
\enar
where the observed angle $n^\alpha$ is defined in a nonperturbative way
and the perturbation order in each quantity can be straightforwardly
understood in conjunction with those in the left-hand side. This relation
explicitly shows that there is only one degree of freedom~$\NC$ and
the wavevector is completely set once the normalization constant is chosen.
By removing the normalization constant at each order, we derive
\bear
\label{eq:fulln}
n^\alpha&=&e^\alpha+\dea^\alpha+\VV^\alpha+\CC_\beta^\alpha e^\beta
+\left\{\dea^\beta+\VV^\beta-{1\over2}e^\gamma\CC_\gamma^\beta
+\left[\dnu+\AA+e^\gamma(\VV_\gamma-\BB_\gamma)\right]e^\beta\right\}
\CC_\beta^\alpha -{1\over2}\left(\VV_\gamma\VV^\alpha
-\BB_\gamma\BB^\alpha\right)e^\gamma \nonumber\\
&&+\bigg[-\dnu-\AA-(\VV_\beta-\BB_\beta) e^\beta
+{3\over2}\AA^2-{1\over2}\VV^\beta\VV_\beta
-3\AA\BB_\beta e^\beta+2\AA\VV_\beta e^\beta
-2\CC_{\beta\gamma}\VV^\beta e^\gamma-\dea^\beta(\VV_\beta-\BB_\beta)
\nonumber\\
&&+\dnu\left[\dnu+\AA+2e^\beta(\VV_\beta-\BB_\beta)\right]
+(\VV_\beta-\BB_\beta)(\VV_\gamma-\BB_\gamma)e^\beta e^\gamma\bigg]e^\alpha
-\left[\dnu+\AA+e^\beta(\VV_\beta-\BB_\beta)\right]\dea^\alpha \nonumber \\
&\equiv&e^\alpha+\dnn^\alpha~,
\enar
where we defined the perturbation~$\dnn^\alpha$ in the observed 
angle~$n^\alpha$ with respect to $e^\alpha$.
Note that the expression is independent of $\NC$, because it is an observable 
quantity, while individual components $\dnu$ and $\dea^\alpha$ are
affected by the choice of the normalization constant and so is $e^\alpha$,
because we split one observable quantity $n^\alpha$ into the mean $e^\alpha$
and the perturbation $\dnn^\alpha$ around it. Furthermore, 
the perturbation is subject to the unit normalization condition:
$n_\alpha n^\alpha=1$, which implies
\beeq
\label{eq:normN}
e_\alpha\dnn^{\alpha(1)}=0~,\quad 2e_\alpha\dnn^{\alpha(2)}+\dnn^{\alpha(1)}
\dnn_\alpha^{(1)}=0~,
\eneq
and the orthogonality condition for another unit directional vector
$\bdi{\that}$ (and similarly for $\bdi{\phat}$)
\beeq
\label{eq:orthN}
e_\theta^\alpha\dnn^{(1)}_\alpha+\dnn_{\ttt\alpha}^{(1)}e^\alpha=0~,\quad
e_\theta^\alpha\dnn^{(2)}_\alpha+\dnn_{\ttt\alpha}^{(2)}e^\alpha
+\dnn^{\alpha(1)}\dnn_{\ttt\alpha}^{(1)}=0~,
\eneq
where the two unit directional vectors constructed from the observed angle 
$\nhat$ are
\beeq
\bdi{\that}={\partial\over\partial\theta}\nhat
\equiv e_\theta^\alpha+\dnn_\ttt^\alpha~,
\quad
\bdi{\phat}={1\over\sin\theta}{\partial\over\partial\phi}\nhat
\equiv e_\phi^\alpha+\dnn_\pp^\alpha~,
\eneq
and a similar set for the unit directional vector~$e^\alpha$ is defined
as $(e_\theta^\alpha, e_\phi^\alpha)$.

\subsection{Observed angle}
\label{app:angle}
The photon wavevector is measured by the observer, and the observed direction
of the source galaxies is independent of our choice of the normalization
constant (or the parametrization of the photon wavevector). However,
the observed direction is characterized by the observed angle 
$(\ttt,\pp)$ of unit directional vector $\nhat$
in the local Lorentz frame as
\beeq
\nhat=n^i_L=(\sin\ttt\cos\pp,~\sin\ttt\sin\pp,~\cos\ttt)~,
\eneq
and these components depend on the choice of frame. For example, suppose
the observer frame is moving with velocity $\bdi{v}$ relative to the rest
frame (say, the rest frame of CMB).
The observed angle in the rest frame is \cite{CHLE02}
\beeq
\label{app:cmb}
\nhat'=\left({\nhat\cdot\vhat-v\over1-\nhat\cdot\bdi{v}}\right)\vhat
+{\nhat-(\nhat\cdot\vhat)\vhat\over\gamma(1-\nhat\cdot\bdi{v})}=
(\sin\theta'\cos\phi',~\sin\theta'\sin\phi',~\cos\theta')~,
\eneq
where $v=|\bdi{v}|$ and $\gamma=(1-v^2)^{-1/2}$.
The aberration due to the relative velocity affects the observed angle.
The generalization of this relation to the general relativistic case
is the four vector of the photon
direction \cite{SASAK87}
\beeq
\label{eq:opho}
\NN^a=-{k^a\over2\pi\nu}+\UU^a~,
\eneq
which satisfies the spacelike condition $\NN^a \NN_a=1$ and the
orthogonality condition $\NN^a\UU_a=0$. Equation~(\ref{app:cmb}) can be
readily derived from Eq.~(\ref{eq:opho}) by Lorentz boosting the observer
velocity.
However, since we are interested in expressing quantities in the FRW frame
in terms of local observable $(\ttt,\pp)$, we have to use the relation
in Eq.~(\ref{eq:fulln}) 
between the local and the FRW components of the photon direction 
expressed in each frame. We use the observed photon direction in 
Eq.~(\ref{eq:opho}) in the FRW frame for computing the fluctuation in the
luminosity distance.

Though Eq.~(\ref{eq:fulln}) is completely general, a dramatic simplification
can be made by a choice of the normalization constant~$\NC$:
While for illustration the normalization constant was split 
into the mean $\bar\NC$ and its fluctuation $\DNC$
in Eq.~(\ref{eq:NCsplit}), the normalization constant represents only {\it one}
degree-of-freedom, and it needs to be specified independent of whether the
universe is homogeneous or inhomogeneous. Our choice of the mean part
in Eq.~(\ref{eq:bgcon}) is automatically
 related to our choice of the perturbation part:
\beeq
\label{eq:concon}
2\pi\hat\nu\equiv-\CG_{ab}\CK^a\CU^b=
2\pi\NC a\nu\bigg|_{\cc_o}\equiv1~,\qquad 
(1+\DNC)(1+\dpnu)\bigg|_{\cc_o}\equiv1~.
\eneq
This condition constrains the perturbations of the photon wavevectors
in Eqs.~(\ref{eq:dnuC1})$-$(\ref{eq:deaC2}) as
\bear
\label{eq:dnuC1o}
\dnu^{(1)}&=&-\AA-n^\alpha(\VV_\alpha-\BB_\alpha)~,\qquad
\dea^{\alpha(1)}=-\VV^\alpha-n^\beta\CC_\beta^\alpha~,\\
\label{eq:dnuC2o}
\dnu^{(2)}&=&-\AA-n^\beta(\VV_\beta-\BB_\beta)
+{3\over2}\AA^2+{1\over2}\VV^\beta\VV_\beta-\VV^\beta\BB_\beta
-n^\beta\left[2\AA\BB_\beta-\AA\VV_\beta+\CC_{\beta\gamma}\left(\VV^\gamma+
\BB^\gamma\right)\right]~,\\
\label{eq:deaC2o}
\dea^{\alpha(2)}&=&-\VV^\alpha-n^\beta\CC_\beta^\alpha
+n^\gamma\left[{1\over2}\left(\VV_\gamma\VV^\alpha
-\BB_\gamma\BB^\alpha\right)+{3\over2}\CC^\beta_\gamma\CC^\alpha_\beta\right]
~,
\enar
where metric perturbations are all evaluated at the observer position
$x^a(\cc_o)$. Using Eq.~(\ref{eq:fulln}), we derive
\beeq
\label{eq:simple}
e^\alpha=n^\alpha~,\quad \dnn^\alpha=0~,
\eneq
the two unit directional vectors of which are described by
the same observed angle ($\ttt,\pp$). Consequently, the other two
orthonormal directional vectors coincide with each other, given
this choice of normalization
condition:
\beeq
\label{eq:simple2}
\nhat\equiv e^\alpha~,\quad
\thatv={\partial\over\partial\theta}\nhat\equiv e_\theta^\alpha~,\quad
\phatv={1\over\sin\theta}{\partial\over\partial\phi}\nhat\equiv e_\phi^\alpha~,
\quad \dnn_\ttt^\alpha=\dnn_\pp^\alpha=0~.
\eneq
Equation~(\ref{eq:cpfa}) (also Eq.~[\ref{eq:fwaves}]) shows that metric
perturbations are well separated from the observable quantities (the
observed angle $n^\alpha$ and the photon frequency~$\nu$), such that
when ensemble averaged given the observable quantities it yields the
desired relations in Eqs.~(\ref{eq:dnuC1o})$-$(\ref{eq:simple2}).
Physically, the normalization condition is a mathematical choice, and our
choice yields $n^\alpha=e^\alpha$ in a homogeneous universe. Therefore, this
relation should remain valid even in an inhomogeneous universe.
This choice for $\dnu$ and $\dea^\alpha$ is found \cite{JESCHI12,SCJE12a}
to the linear order in perturbations. However, we explicitly provide physical
justification as to the presence of additional degree of freedom
and its relation to the photon wavevector~$k^a$. It is noted that
there is only one degree-of-freedom in the normalization constant 
(i.e., $\bar\NC$ and $\DNC$ are not independent), and
other choice of the normalization constant~$\NC$ is equally valid, while 
other choice would significantly
complicate the relation between $e^\alpha$ and $n^\alpha$.

%% file: table.tex
\begin{table*}
\caption{Symbols in the paper}
\begin{ruledtabular}
\begin{tabular}{ccc}
Symbol & Definition & Equation\\
\hline
$\zz$ & observed redshift of source galaxy & (\ref{eq:obsz})\\
$\ttt$, $\pp$ & observed angular position of source galaxy $\nhat=(\ttt,\pp)$
& $\cdot$ \\
$n^\alpha$ & unit vector in FRW coordinate for observed angle $\nhat$
& (\ref{eq:fulln})\\
$n_g$, $n_g^\up{obs}$ & physical and observed galaxy number densities & 
(\ref{eq:ngobs1}) and (\ref{eq:ngobs2}) \\
$dV_\up{phy}$, $dV_\up{obs}$ & physical and observationally inferred
volumes occupied by source galaxies & 
(\ref{eq:ddv}) and (\ref{eq:obsdv}) \\
$\dV$ & distortion in volume between $dV_\up{phy}$ and $dV_\up{obs}$ & 
(\ref{eq:ddv})$-$(\ref{eq:ddv2}) \\
$e_1$, $e_2$ & coefficients describing the time
evolution of galaxy sample & (\ref{eq:eee}) \\
$t_1$, $t_2$ & coefficients describing the luminosity function of 
galaxy sample & (\ref{eq:ttt}) \\
$\delta_g^\up{int}$, $\delta_g^\up{obs}$ & intrinsic fluctuation of
galaxy number density, observationally constructed galaxy fluctuation & 
(\ref{eq:bng}) and (\ref{eq:dobs0})$-$(\ref{eq:dobs2})\\
\hline
$\hat x^a_s$ & observationally inferred position of source galaxies 
& (\ref{eq:inferred}) \\
$x^a_s=(\tau_s,x^\alpha_s)$ 
& real source galaxy position & (\ref{eq:srcone})$-$(\ref{eq:srctwo2})
and (\ref{eq:displ}) \\
$\bar x^a_s=(\bar\tau_s,\bar x^\alpha_s)$ 
& source galaxy position in a homogeneous universe & 
(\ref{eq:srczero}) \\
$\dz$ & distortion in observed redshift & (\ref{eq:obsz})\\
$\drr$ & distortion in comoving distance from the inferred distance
 based on observed redshift & (\ref{eq:displ})\\
$\dtt$, $\dpp$ & distortion in angle from observed angle & (\ref{eq:displ})\\
$\DT_z(=\dTau)$, $\DX_z^\alpha$  & distortion in comoving coordinate
from the inferred & (\ref{eq:DT}) and (\ref{eq:DX})\\
\hline
$k^a$ & photon wavevector & (\ref{eq:fwave})\\
$\CK^a$ & conformally transformed photon wavevector & (\ref{eq:conf}) and
(\ref{eq:cpf})$-$(\ref{eq:cpfa})  \\
$k_L^a$ & photon wavevector in the observer rest frame & (\ref{eq:lwave}) \\
$\nu$, $\omega$ & frequency and angular frequency of photon & 
(\ref{eq:lwave}) \\
$[e_\mu]^a$ & orthnormal tetrads vectors & (\ref{eq:tet0})$-$(\ref{eq:tet2})\\
$\bar\nu$ & photon frequency in a homogeneous universe & (\ref{eq:bgcon}) \\
$\dhnu$ & perturbation part in $\CK^a\CU_a$ & (\ref{eq:cnu})
and (\ref{eq:concon}) \\
$\oo$ & physical affine parameter & (\ref{eq:affineoo})\\
$\cc$ & conformally transformed affine parameter & (\ref{eq:conf}) and
(\ref{eq:affine}) \\
$\Dcc_s$ & perturbation in conformally transformed affine parameter
 & (\ref{eq:matchz})\\
$\NC$ & normalization constant due to conformal transformation
 & (\ref{eq:conf}) and (\ref{eq:concon}) \\
$\dnu$, $\dea^\alpha$ & perturbations in temporal and spatial components of 
the photon wavevector & (\ref{eq:wave}) \\
$\eP$, $\eTT$, $\ePP$ & spatial decomposition of $\dea^\alpha$
along ($\nhat$, $\thatv$, $\phatv$) & (\ref{eq:deadec}) \\
$\NN^a$ & observed photon four vector in FRW coordinate & (\ref{eq:opho})\\
\hline
$\UU^a$ & four velocity & (\ref{eq:fourv}) \\
$\dUU^0$ & perturbation in temporal component of conformally transformed
four velocity & (\ref{eq:fourv})\\
$\VV^\alpha$ & spatial component of conformally transformed four velocity
 & (\ref{eq:fourv})\\
$\VP$, $\VTT$, $\VPP$ & decomposition of spatial velocity along ($\nhat$,
$\thatv$, $\phatv$) & (\ref{eq:vvdec}) \\
$\kappa$ & gravitational lensing convergence & (\ref{eq:kappadef}) \\
$\dL$, $\mathcal{D}_A$ & physical luminosity and angular distances & 
(\ref{eq:dL}) and (\ref{eq:dA}) \\
$\DD$ & deformation matrix in angle & (\ref{eq:deform}) \\
$\dDD$ & second-order distortion in solid angle & (\ref{eq:Ddet})\\
$\ddL$ & fluctuation in luminosity distance & (\ref{eq:dL})\\
\hline
$\chi$ & spatially gauge-invariant metric perturbation variable 
& (\ref{eq:chid})\\
$\px$ & scalar gauge-invariant variable in metric tensor & (\ref{app:pxgi})\\
$\ax$ & scalar gauge-invariant variable in metric tensor & (\ref{app:etcgi})\\
$\dnug$ & temporal gauge-invariant variable for photon wavevector 
& (\ref{app:dnug})\\
$\deag$ & spatial gauge-invariant variable for photon wavevector
& (\ref{app:deag})\\
$\pv_{\chi\alpha}$ & vector gauge-invariant variable in metric tensor
& (\ref{app:pv})\\
$\TCC_{\chi\alpha\beta}$ & tensor gauge-invariant variable in metric tensor
& (\ref{app:tcc})\\
$\drr_\chi$, $\dtt_\chi$ 
& gauge-invariant radial and angular distortions of the source galaxy position
 & (\ref{eq:drg})$-$(\ref{eq:dtg}) \\
\end{tabular}
\end{ruledtabular}
\label{tab:symbol}
\end{table*}

\clearpage
\begin{table*}
\caption{Symbols in the paper (continued)}
\begin{ruledtabular}
\begin{tabular}{ccc}
Symbol & Definition & Equation\\
\hline
$a$ & comoving scale factor & (\ref{eq:bgmetric})\\
$H$, $\HH$ & Hubble and conformal Hubble parameters & $\cdot$\\
$\rbar$ & comoving line-of-sight distance & (\ref{eq:affine}) \\
$g_{ab}$, $\CG_{ab}$ & metric and conformally transformed metric
tensors & (\ref{eq:abc}) and (\ref{eq:conf}) \\
$g$, $\dg$ & metric determinant and perturbation part & (\ref{eq:ggg}) \\
$\gbar_{\alpha\beta}$ & metric tensor for a three-space in a homogeneous
universe & (\ref{eq:bgmetric}) \\
$\eta_{\mu\nu}$ & Minkowsky metric in a local frame & $\cdot$ \\
$\AA$ & temporal perturbation in metric tensor & (\ref{eq:abc})\\
$\BB_\alpha$ & off-diagonal perturbation in metric tensor & (\ref{eq:abc})\\
$\CC_{\alpha\beta}$ & spatial perturbation in metric tensor & (\ref{eq:abc})\\
$\alpha$ & temporal perturbation in metric tensor & (\ref{eq:decmet})\\
$\beta$ & off-diagonal scalar perturbation in metric tensor 
& (\ref{eq:fourv})\\
$\VBB_\alpha$ & off-diagonal vector perturbation in metric tensor 
& (\ref{eq:fourv})\\
$\varphi$ & scalar spatial perturbation in metric tensor 
& (\ref{eq:fourv})\\
$\VCC_\alpha$ & spatial vector perturbation in metric tensor 
& (\ref{eq:fourv})\\
$\TCC_{\alpha\beta}$ & spatial tensor perturbation in metric tensor 
& (\ref{eq:fourv})\\
$\xi^a=(T,\mathcal{L}^\alpha)$ & coordinate transformation & (\ref{eq:gt}) 
and (\ref{eq:gtdec})\\
$L$, $L^\alpha$ & scalar and vector decomposition of $\mathcal{L}^\alpha$
& (\ref{eq:gtdec}) \\
$\GG^0$, $\GG^\alpha$ & source perturbations in photon geodesic equation
& (\ref{eq:GG0}) and (\ref{eq:GGA})
\end{tabular}
\end{ruledtabular}
\end{table*}